\documentclass[11pt]{article}
\usepackage[final]{epsfig}
\usepackage{latexsym}
\usepackage{graphicx}
\usepackage{subfigure}
\usepackage{float}
\usepackage{wrapfig}
\usepackage{indentfirst}
\usepackage{xcolor}
\usepackage{amssymb,amsmath,amsfonts,latexsym,amsthm}
\usepackage[font=small,labelfont=bf]{caption}
\usepackage{fancyhdr,hyperref}
\usepackage[small,compact]{titlesec}
\usepackage{url}
\usepackage{authblk}
\usepackage{microtype}
\usepackage[T1]{fontenc}
\usepackage[latin9]{inputenc}
\usepackage{geometry}
\usepackage{verbatim}
\usepackage{graphicx}
\PassOptionsToPackage{normalem}{ulem}
\usepackage{ulem}
\usepackage{natbib}

\newcommand{\angstrom}{\mbox{\normalfont\AA}}

\newcommand{\bfe}{{\bf e}}

\newcommand{\bfI}{{\bf I}}

\newcommand{\beq}{\begin{equation}}
\newcommand{\eeq}{\end{equation}}
\newcommand{\beqs}{\begin{eqnarray}}
\newcommand{\eeqs}{\end{eqnarray}}

\title{\textcolor{black}{Film strains} enhance the reversible cycling of intercalation electrodes}
\author[1]{Delin Zhang}
\author[2]{Jay Sheth}
\author[2]{Brian W. Sheldon}
\author[1*]{Ananya Renuka Balakrishna}
\affil[1]{\small{Aerospace and Mechanical Engineering, University of Southern California, Los Angeles, CA 90089}}
\affil[2]{\small{School of Engineering, Brown University, Providence, RI 02912}}
\affil[*]{\small{Corresponding author, Office: (213) 740-8762, Email: renukaba@usc.edu}}
\date{}
\usepackage{babel}

\begin{document}
\maketitle

\begin{abstract}
A key cause of chemo-mechanical degradation in battery electrodes is that they undergo abrupt phase transformation during the charging/discharging cycle. This phase transformation is accompanied by lattice misfit strains that nucleate microcracks, induce fracture and, in extreme cases, amorphize the intercalation electrode. In this work, we propose a strategy to prevent the chemo-mechanical degradation of intercalation electrodes: we show that by engineering suitable film strains we can regulate the phase transformations in thin-film intercalation electrodes and circumvent the large volume changes. We test this strategy using a combination of theory and experiment: we first analytically derive the effect of film strain on the electrochemical response of a thin-film intercalation electrode and next apply our analytical model to a representative example ($\mathrm{Li_xV_2O_5}$ with multiple phase transformations). We then test our theoretical predictions experimentally. Specifically, we electrochemically cycle thin-film $\mathrm{V_2O_5}$ electrodes with different film strains and measure their structure, voltage, and stress responses. Our findings show that tensile film strains lower the voltage for phase transformations in thin-film $\mathrm{V_2O_5}$ electrodes and facilitate their reversible cycling across a wider voltage window without chemo-mechanical degradation. These results suggest that film strain engineering is an alternative approach to preventing chemo-mechanical degradation in intercalation electrodes. Beyond thin-film electrodes, our findings from this study are applicable to the study of stress-induced phase transformations in particle-based electrodes and the thin surface layers forming on cathode particles. 
\end{abstract}

\section{Introduction}
Lithium-ion batteries (LIBs) are promising candidates for sustainable energy storage, although their use in high-energy-density applications, such as powering electric vehicles and aircrafts, is still a challenge. One of the key limitations for battery performance is the structural degradation of electrodes with repeated cycling, ultimately resulting in capacity loss \citep{radin2017narrowing,lewis2019chemo,christensen2018structural,mukhopadhyay2014deformation}. When charging/discharging a battery, most intercalation cathode materials undergo volume changes of $\sim10-15\%$ \citep{xiang2017accommodating}. These are particularly problematic when they are associated with abrupt lattice transformations that induce misfit strains \citep{poluektov2019micromechanical}. Upon repeated cycling, the corresponding stresses can nucleate microcracks and in some cases amorphize the materials \citep{chen2006electron,shpigel2019diffusion,zhang2018recent}. These mechanisms severely limit the reversible cycling of a variety of electrodes and contribute to the structural degradation of batteries \citep{de2018striping,
zhang2019chemo}.

In the past, researchers have developed several approaches to limit the structural degradation of intercalation materials. For example, operating a vanadium pentaoxide ($\mathrm{V_2O_5}$) electrode in a smaller voltage window 2.5--4V can prevent the large volume changes at $\gamma-\omega$ transformation \citep{shimizu1993electrochemical,baddour2012structural,whittingham1976role}. In other intercalation electrodes like $\mathrm{FePO_4}$, phase transformation is controlled by regulating the Li-diffusion kinetics  \citep{woodford2012design,kao2010overpotential,tang2009model,bai2011suppression} and/or by fabricating electrodes as nanoparticles \citep{liu2011v2o5,yue2017micro,wang2006nanostructured}. These experiments have resulted in a solid-solution type of microstructure that does not contain phase boundaries and thus avoids abrupt changes in elastic strains. While these strategies avoid large transformation strains in intercalation electrodes and have improved their lifespans, they do so at the cost of the material's energy storage capacity and performance \citep{radin2017narrowing,mukhopadhyay2014deformation}.

Alternatively, we may look into other materials systems, such as semiconductors, ferroelectrics, and dichalcogenides, in which phase transformations are regulated by film strains \citep{sun2017substrate,roytburd2015reversible,ota2016strain,graz2009extended}. In thin-film ferroelectrics $\mathrm{SrTiO_3}$, the misfit strains are precisely controlled to increase the phase transformation temperature from cryogenic values ($\approx\mathrm{0K}$) to room temperature \citep{haeni2004room,chen2008phase}. The phase transformation microstructures such as periodic arrangement of metal-insulator phases in $\mathrm{VO_2}$ are modulated using film strains \citep{wu2006strain,shao2018recent}. Additionally, in thin-film Co electrodes tensile stresses are used to delay the crack onset  during phase transformations \citep{marx2016strain}. However, for intercalation electrodes such as $\mathrm{V_2O_5}$, little is known about the extent to which engineering film strains can alter the phase transformation voltages. Doing so could expand the operational voltage window for $\mathrm{V_2O_5}$ electrodes while retaining the structural integrity of electrodes.

Today, most practical batteries use a particle-based electrode with the intercalation material as its active particle \citep{mohanty2016modification,crabtree2015energy}. In these electrode configurations, it is challenging to investigate the effect of mechanical stresses on phase changes and voltage curves in the intercalation material. This is largely because of the difficulty in controlling and assessing the mechanical constraints and the corresponding stress states that occur in particle-based electrodes. These issues are generally easier to evaluate with thin-film electrode configurations because of their large surface area and relatively easy-to-control morphology. Thin-film electrode configurations are used in microbatteries \citep{fehse2017ultrafast,nathan2005three}, and resemble the thin surface layers that form on many cathode particles (e.g., in $\mathrm{LiNi_{1-y-z}Co_{y}Mn_{z}O_2}$ (NMC) cathodes, where phase transformations create surface damage layers that lead to capacity loss \citep{sallis2016surface}). Previously, \cite{bucci2014measurement} have investigated the electrochemical-mechanical response of thin-film amorphous Si electrodes, but the links between film strains and structural degradation of crystalline electrodes is not well understood. With this in mind, for this study, we propose to investigate the role of film strains on phase transformations in thin-film intercalation electrode geometry. To be specific, we will evaluate thin film configurations where the volume changes of the electrode are constrained by an underlying stiff substrate and investigate the role of film strains on phase transformation voltages in intercalation electrodes.

We hypothesize that the film strains can regulate the voltages for phase transformations in electrode materials, and thus can improve the reversible cycling over a wider voltage range. We test this hypothesis using a combination of theory and experiments: First, we analytically derive the effect of film strain on the electrochemical response of cathodes. Our analytical framework is general and in principle can be adapted to any intercalation material. As an example for the present study, we choose the $\mathrm{V_2O_5}$ compound. $\mathrm{V_2O_5}$ is a relevant intercalation electrode for Li-batteries and can store up to three Li-ions per formula unit via a series of phase transformations.\footnote{The phases of interest include the $\alpha, \epsilon, \delta, \gamma$  and $\omega$ phases, which respectively correspond to Li-contents, $x<0.01$,  $0.35<x<0.7$, $x=1$, $1<x<3$, $x=3$.} These phase transformations occur over a wide range of voltage windows \citep{wang2006synthesis,christensen2018structural} and are accompanied by substantial volume changes in $\mathrm{V_2O_5}$.\footnote{For example, the $\gamma-\omega$ $(x\approx3)$ transformation is particularly problematic because it leads to irreversible structural damage and amorphization of the material \citep{christensen2018structural,huang2019high}. Consequently, although vanadium pentaoxide ($\mathrm{V_{2}O_{5}}$) promises high power densities and high capacities \citep{jeon2001characterization,wang2006synthesis}, its structural irreversibility limits its widespread usage.} We theoretically investigate whether thin-film straining of $\mathrm{Li_xV_2O_5}$ electrode can regulate its multiple phase transformation stages and thus enable the structural reversibility of the intercalation material. Second, we test our theoretical predictions using experiments. Here, we electrochemically cycle two types of thin-film $\mathrm{V_2O_5}$ electrodes with different film strains and investigate their corresponding mechanical degradation. Doing so enables us to measure the in-situ stress evolution in electrodes during electrochemical cycling. Our experimental findings validate our theoretical predictions that film strains modify the voltage for phase transformation and help improve the structural reversibility of intercalation cathodes. Overall, our work presents a theoretical and experimental framework to engineer film strains as design criteria to enhance the lifespans of intercalation materials.

\section{Theory}
In this section we derive the effect of film strain on phase transformations in thin-film intercalation electrodes, \textcolor{black}{and show how film strains can be engineered to enhance the structural reversibility and lifespans of electrodes}. First, we describe the free energy function for an intercalation electrode and analytically derive the effect of film strain on the free energy landscape. Next, we investigate how film strain affects the chemical potential and stress evolution, in-situ, in thin-film electrodes. 

In principle, our theoretical framework can be adapted to any phase separating electrode material, but for the present work, we choose $\mathrm{Li_xV_2O_5}$ as the reference electrode. Crystalline $\mathrm{Li_xV_2O_5}$ undergoes series of phase transformations as a function of Li-content and each phase transformation stage is characterized by a constant voltage plateau describing a two-phase coexistence \citep{christensen2018structural}. In our analytical calculations we model such two-phase coexistence and investigate the effect of elastic energy on individual voltage plateaus. For further details on the phase transformations in $\mathrm{Li_xV_2O_5}$ and its corresponding material parameters see Section \ref{Theoretical calculations} of the Appendix.

\subsection{Free energy}
The free energy $\Psi$ of an electrode $\mathcal{E}$ as a function of Li-content $c$ and strain $\mathbf{E}$ is given by:

\begin{align}
\Psi(c,\nabla c,\mathbf{E}) & = \psi_{grad}(\nabla c) + \psi_{bulk}(c) + \psi_{elas}(\mathbf{E},c)\nonumber\\
 & =\int_{\mathcal{E}}\frac{1}{2}\kappa|\nabla c|^{2}+\frac{\Omega}{V_m}c(1-c)+\frac{RT}{V_m}[c\mathrm{ln}(c)+(1-c)\mathrm{ln}(1-c)]\nonumber \\
 & +\frac{1}{2}[\mathbf{E-E}_{0}(c)]\textcolor{black}{\ :\ }\mathbb{C}[\mathbf{E-E}_{0}(c)]\thinspace\mathrm{d}\mathbf{x}\label{eq:Free Energy}
\end{align}

The total free energy of a heterogeneous system given by Eq.~\ref{eq:Free Energy} is the volume integral of the gradient energy, bulk energy and the elastic energy of the system. The Li-content field $c(\mathbf{x})$ represents the fraction of interstitial sites in the electrode occupied by Li per unit volume. The gradient energy $\int\kappa|\nabla c|^{2}\mathrm{d\mathbf{x}}$ penalizes changes in Li-content  and its coefficient $\kappa$ is determined in numerics/computations to define the interface (e.g., phase boundaries) properties. This coefficient is typically small and its contribution to the total energy at larger length scales is often neglected. The bulk energy (a regular solution model) describes the free energy landscape with minima corresponding to the lithiated $(c=1)$ and unlithiated $(c=0)$ phases of the electrode. The regular solution coefficients are calibrated to fit the thermodynamic properties to voltage measurements of $\mathrm{Li_xV_2O_5}$. The elastic energy of the system accounts for \textcolor{black}{elastic strains arising from lattice misfit between the lithiated/unlithiated phases and the film strains arising from the mismatch between the film and substrate. These strains deviate from the preferred strain tensor $\mathbf{E}_{0}(c)$} that describes the transformation of lattice geometry between the lithiated and unlithiated phases. In this paper, we assume $\mathbf{E}_{0}(c)$ to vary as a function of Li-content $c$ and is given by:

\begin{align}
\mathbf{E}_{0}(c) & =\left[\begin{array}{ccc}
\epsilon_{\alpha}(c) & 0 & 0\\
0 & \epsilon_{\beta}(c) & 0\\
0 & 0 & \epsilon_{\gamma}(c)
\end{array}\right].\label{eq:PreferredStrainTensor}
\end{align}

Here, $\epsilon_{\alpha}(c), \epsilon_{\beta}(c), \epsilon_{\gamma}(c)$ are polynomials of third-order in $c$ that are fitted to the lattice parameter measurements of $\mathrm{Li_xV_2O_5}$, and are listed in Table.~\ref{table:coefficients} of Appendix. These polynomials account for the anisotropic lattice strains in the intercalation material, which differs from the elastic strains defined for alloying-type of electrodes \citep{sheldon2011stress}. From here on we note that the preferred strain tensor is a function of Li-content, and we drop $c$ from the notation. Similarly the polynomials of the preferred strains will be represented by  $\epsilon_{\alpha}, \epsilon_{\beta}$ and $\epsilon_{\gamma}$.

\subsection{\textcolor{black}{Film strain} modifies the free energy landscape}

\begin{figure}[H]
\centering
\includegraphics[width=\textwidth]{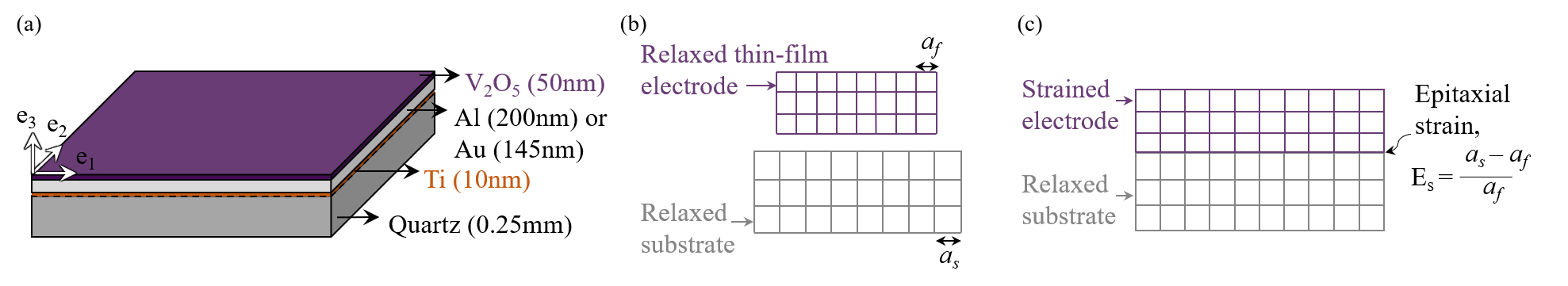}
\caption{\label{fig:Schematic-illustration-of-thin-film}(a) Schematic illustration of a thin-film electrode (e.g., $\mathrm{V_{2}O_{5}}$) constrained to a much thicker substrate. The substrate typically consists of two layers, namely the current collectors (e.g., aluminum (Al) or gold (Au)) and a thick quartz substrate. An additional titanium layer (Ti) is used with the gold collector for better adhesivity. (b) The film strains arise from the CTE mismatch between the thin film and the substrate. For example, after annealing the thin-film shrinks more than the quartz substrate. (c) When the film is bound to the substrate the CTE mismatch induces a film strain in the electrode. }
\end{figure}

Next, we introduce the boundary conditions for a thin-film electrode in Eq. \ref{eq:Free Energy}. A thin-film of a single-crystal electrode constrained to a rigid substrate is shown in Fig. \ref{fig:Schematic-illustration-of-thin-film}, where the mismatch between the substrate and the electrode lattices imposes a film strain $\mathrm{E_{s}}$ along the $\mathrm{e_{1}-e_{2}}$ directions. \textcolor{black}{Such mismatch could arise from differences in coefficient of thermal expansions (CTE) between the film and substrate.} For the thin-film shown in Fig.~\ref{fig:Schematic-illustration-of-thin-film}, we model inplane axial strains $\mathrm{E_{11}=E_{22}=E_{s}}$ and $\mathrm{E_{12}=E_{21}=0}$. We calculate the film strain $\mathrm{E_{s}}$ as a CTE mismatch between the unlithiated film and the substrate material.

In thin-film electrodes, the film thickness is far smaller than its in-plane dimensions and is acted only upon by loads (e.g., CTE misfit) that are parallel to the film. With this electrode geometry and boundary conditions, we assume that thin-films are under the plane stress condition, and thus satisfy the stress-free boundary conditions along the $\mathrm{e_{3}}-$axis. That is, the stresses involving the out-of-plane components are zero, $\sigma_{13}=\sigma_{23}=\sigma_{31}=\sigma_{32}=0.$
Consequently, the strains $\mathrm{E_{13}, E_{23}, E_{31}, E_{32}}$
corresponding to their respective stress components are zero. The strain tensor for a thin film reduces to:

\begin{align}
\mathbf{E} & =\left[\begin{array}{ccc}
\mathrm{E_{s}} & 0 & 0\\
0 & \mathrm{E_{s}} & 0\\
0 & 0 & \mathrm{E_{33}}
\end{array}\right].\label{eq:StrainTensorEpitaxialThinFilm}
\end{align}

During lithiation and delithiation, the thin-film deforms
in the out-of-plane direction to minimize the total elastic energy of the system. In other words, the out-of-plane strain $\mathrm{E_{33}}$ satisfies the plane stress condition $\sigma_{33}=0$ and is obtained by minimizing the free
energy with respect to $\mathrm{E_{33}}.$ 

\begin{align}
\frac{\delta\Psi}{\delta\mathrm{E_{33}}} & =\mathrm{C_{11}}[\mathrm{E_{33}}-\epsilon_{\gamma}]+\mathrm{C_{12}}[2\mathrm{E_{s}}-\epsilon_{\beta}-\epsilon_{\alpha}]\label{eq:MinimizingPsi}
\end{align}

Next, in order to satisfy $\sigma_{33}=0,$ we have:

\begin{align}
\frac{\delta\Psi}{\delta\mathrm{E_{33}}} & =0\nonumber \\
\mathrm{E_{33}} & =\mathrm{\frac{C_{12}}{C_{11}}}[\epsilon_{\alpha}+\epsilon_{\beta}-2\mathrm{E_{s}}]+\epsilon_{\gamma}\label{eq:E33}
\end{align}

The out-of-plane strain $\mathrm{E_{33}}$ is a function of the film strain and Li-content (i.e., $\epsilon_\alpha, \epsilon_\beta, \epsilon_\gamma$ are polynomials in $c$). Substituting Eq. \ref{eq:E33} in Eq. \ref{eq:Free Energy}, and assuming a linear elastic cubic crystal\footnote{The indicial form of the elastic energy for a linear cubic crystal is derived in Section \ref{Theoretical calculations} of Appendix.} we can express the free energy of the electrodes as a function of Li-content $c$ and film strain $\mathrm{E_{s}}$ as follows:

\begin{align}
\Psi(c,E_{\mathit{s}})=\int_{\mathcal{E}}\bigg\{&\frac{1}{2}\kappa|\nabla c|^{2}+\frac{\Omega}{V_m}c(1-c)+\frac{RT}{V_m}[c\mathrm{ln}(c)+(1-c)\mathrm{ln}(1-c)]\nonumber\\
 & +\mathrm{\frac{C_{11}-C_{12}}{2}}\bigg[(\mathrm{E_s}-\epsilon_{\alpha})^2+(\mathrm{E_s}-\epsilon_{\beta}])^{2} + \bigg(\mathrm{\frac{C_{12}}{C_{11}}}[\epsilon_{\alpha}+\epsilon_{\beta}-2\mathrm{E_{s}}]\bigg)^{2}\bigg]\nonumber \\
& +\frac{\mathrm{C_{12}}}{2}[\mathrm{E_s}-\epsilon_{\alpha} + \mathrm{E_s}-\epsilon_{\beta} + \mathrm{\frac{C_{12}}{C_{11}}}(\epsilon_{\alpha}+\epsilon_{\beta}-2\mathrm{E_{s}})]^2\bigg\}\mathrm{d}\mathbf{x}\label{eq:EnergyLandscape}
\end{align}

\begin{figure}[H]
\centering
\includegraphics[width=0.8\textwidth]{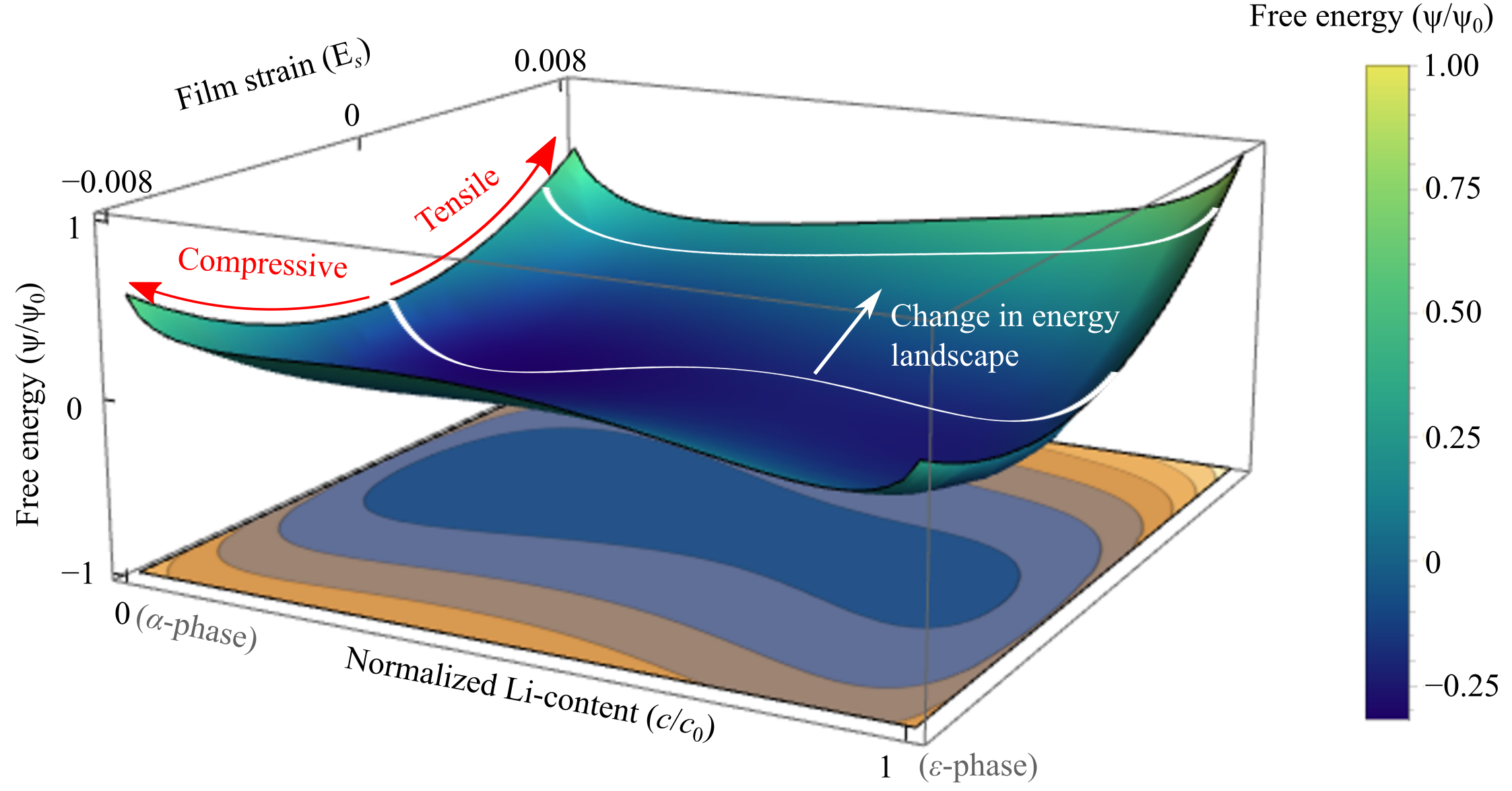}
\caption{A 3D plot of the free energy expression in Eq.~\ref{eq:EnergyLandscape} as a function of Li-content and film strain. The coefficients correspond to the $\alpha-\epsilon$ phase transformation in $\mathrm{Li_xV_2O_5}$ electrode. \textcolor{black}{Film strains affect the height of wells and alter the energy barrier between two phases.} As can be seen, increasing tensile strains shifts the height of minimum energy wells and lowers the energy barrier between the $\alpha$ and $\epsilon$ phases.}
\label{fig:energy-landscape}
\end{figure}

The coefficients in Eq.~\ref{eq:EnergyLandscape} correspond to the $\alpha-\epsilon$ phase transformation stage in $\mathrm{V_2O_5}$ and are listed in Appendix. We next plot Eq.~\ref{eq:EnergyLandscape} as a function of $c$ and $\mathrm{E_{s}}$ (see Fig.~\ref{fig:energy-landscape}). At $\mathrm{E_s} = 0$, a double-well landscape with minima located at the unlithiated $c/c\mathrm{_0} = 0$ and the lithiated phases $c/c\mathrm{_0} = 1$ is shown. Here $c_\mathrm{0}$ denotes the Li-mole fraction of the $\epsilon$ phase. This double-well landscape changes in two ways under non-zero film strains: First, increasing tensile (or compressive) strains shift the minimum energy wells. \textcolor{black}{For example, the height of the energy wells changes with increasing tensile strain, and at $\mathrm{E_s = 0.008}$ the global minima is located at $c=0$.} Second, increasing tensile (or compressive) strains affects the height of the energy barrier between the two wells. That is, under tensile strains the energy barrier between $c=0$ and $c=c_\mathrm{0}$ wells is reduced and thus would require a smaller driving force for phase transformation. In the next subsection, we will show how these changes to the free energy landscape can be engineered to facilitate reversible phase transformations in battery electrodes.

\subsection{Chemical potential}
We define the chemical potential as the variational derivative of the free energy functional in Eq.~\ref{eq:Free Energy}:
\begin{align}
\mu & =  \frac{\delta\Psi}{\delta c} \nonumber\\
& = \frac{\partial\psi_{bulk}}{\partial c} + \frac{\partial\psi_{elas}}{\partial c} - \kappa\nabla^2 c \label{eq:ChemicalPotential}
\end{align}

\begin{figure}[H]
\centering
\includegraphics[width=\textwidth]{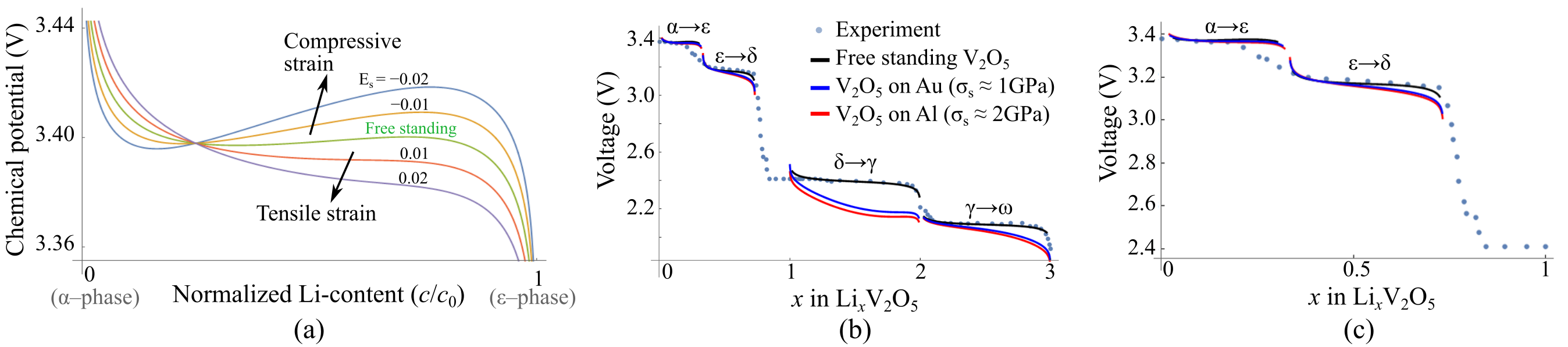}
\caption{\label{fig:chemical potential}(a) The film strain affects the voltage plateaus corresponding to the phase transformation stage. Specifically, increasing tensile strains introduce a downward slope to the chemical potential curve and shift the phase transformation voltage to lower values. (b) Using Eq.~\ref{eq:ChemicalPotential} we predict the chemical potential landscape for each phase transformation stage in $\mathrm{Li_xV_2O_5}$ electrode. We note that in addition to voltage plateaus lowering, tensile strains also introduce an additional slope to the chemical potential curves. (c) Inset figure showing the $\alpha-\epsilon$ and $\epsilon-\delta$ phase transformations in $\mathrm{Li_xV_2O_5}$ electrode. The experimental data corresponding to the electrochemical measurements for a free-standing $\mathrm{V_2O_5}$ electrode is provided by \cite{luo2018roadblocks}.}
\end{figure}

Fig.~\ref{fig:chemical potential}(a) depicts the chemical potential as a function of Li-content in the $\mathrm{V_2O_5}$ intercalation cathode. The flat voltage plateau at $-\mu/\mathrm{F}=3.4\mathrm{V}$ represents phase separation during the $\alpha-\epsilon$ phase transformation for a free-standing $\mathrm{V_2O_5}$ electrode (Note $\mathrm{F = 96485 \thinspace C/mol}$ is the Faraday constant). This flat voltage plateau gains an upward (or downward) slope with increasing compressive (or tensile) film strains. Additionally, the plateau shifts to lower voltages with increasing tensile strains.\footnote{Note that the effect of tensile and compressive strains on the voltage plateau in Fig.~\ref{fig:chemical potential}(a) is specific to the $\alpha-\epsilon$ phase transformation of $\mathrm{V_2O_5}$ and its corresponding anisotropic lattice strains. For example, the lattice strains $\epsilon_\alpha$ decreases by 0.5\% on intercalation while $\epsilon_\beta$ and $\epsilon_\gamma$ increase by 0.1\% and 3.2\% respectively on intercalation. This anisotropic expansion of $\mathrm{Li_xV_2O_5}$ affects the observed shift of the voltage plateau.}

\textcolor{black}{Fig.~\ref{fig:chemical potential}(b-c) shows four distinct voltage plateaus corresponding to four stages of phase transformation of $\mathrm{Li_xV_2O_5}$. Although all four voltage plateaus are lowered with increasing film strains, significant shifts in voltage curves are observed for the $\delta-\gamma$ and the $\gamma-\omega$ phase transformation stages. We attribute these relatively large shifts of voltage plateaus to the abrupt lattice transformation at the $\delta-\gamma$ and the $\gamma-\omega$ phase transformations, which generate lattice strains of up to $\sim10\%$ (see Fig.~\ref{fig:strain-fit} in Appendix). This abrupt change in lattice strains, in particular the $\epsilon_{\alpha}(c)$, affects the chemical potential via the elastic energy term.}

\begin{figure}[H]
\centering
\includegraphics[width=0.7\textwidth]{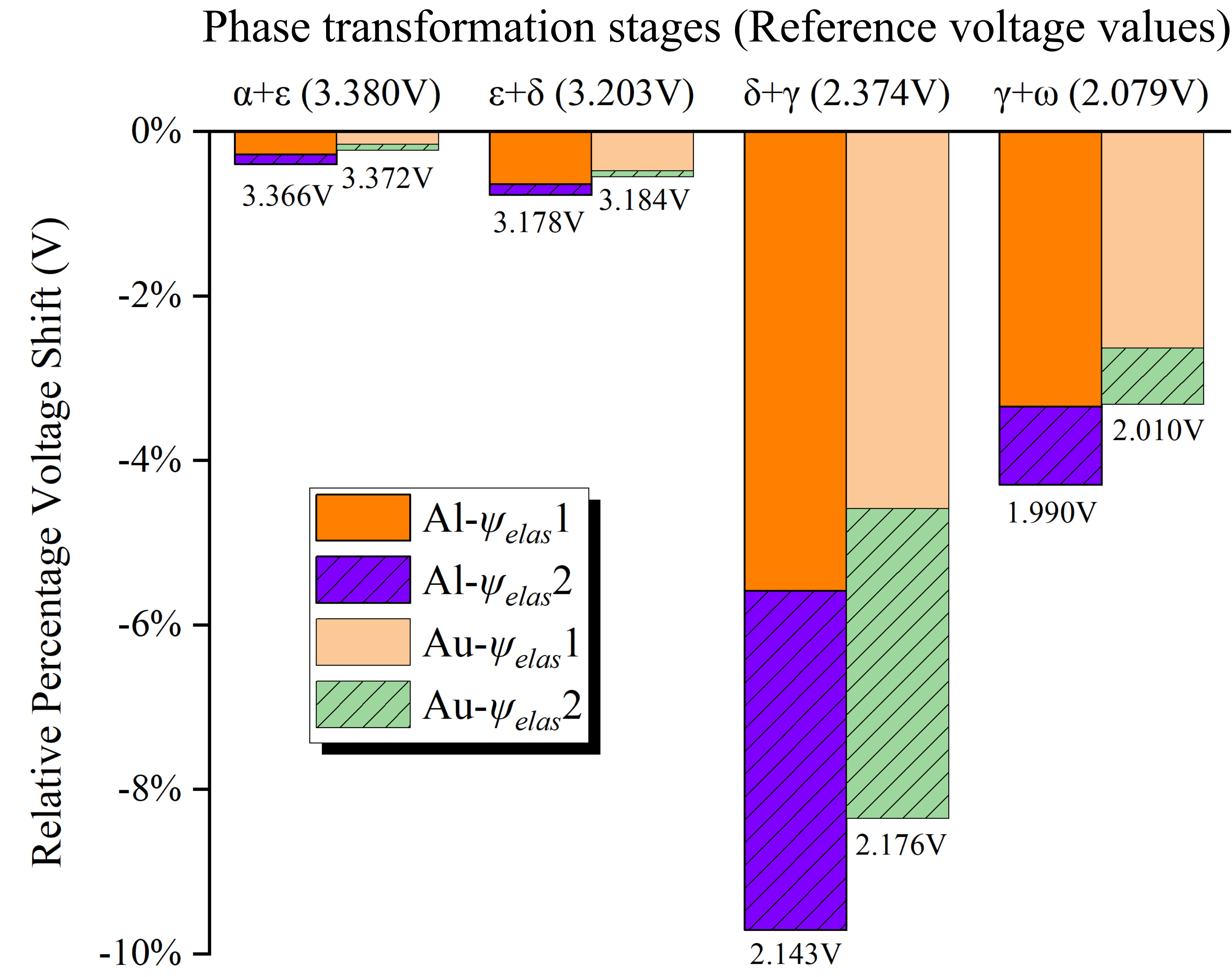}
\caption{\textcolor{black}{The effect of $\psi_{elas}1$ and $\psi_{elas}2$ to the percentage voltage shift. Here relative percentage voltage shift is derived by comparing with reference voltage values of a free-standing $\mathrm{Li_xV_2O_5}$ in each phase transformation stage.}}
\label{fig:percentage shift}
\end{figure}

We next quantify the effect of individual elastic energy terms in Eq.~\ref{eq:EnergyLandscape} on the voltage plateaus shifts. For a free standing $\mathrm{Li_xV_2O_5}$ intercalation electrode the strain tensor will closely satisfy the preferred lattice strains, i.e., $\mathbf{E} = \mathbf{E}_0(c)$ to minimize the elastic energy of the system. However, the thin-film geometry of the electrode imposes a mechanical constraint with $\mathrm{E_{11}} = \mathrm{E_{22}} = \mathrm{E_s}$, which gives rise to a finite elastic energy in the system. Fig.~\ref{fig:percentage shift} shows how the elastic energy terms, namely the terms on lines two ($\psi_{elas}1$) and three ($\psi_{elas}2$) of Eq.~\ref{eq:EnergyLandscape}, affect the voltage plateau shifts.\footnote{$\psi_{elas}1 = \frac{c_{11}-c_{12}}{2}[(\mathrm{E_s}-\epsilon_{\alpha})^2+(\mathrm{E_s}-\epsilon_{\beta}])^{2} + (\frac{c_{12}}{c_{11}}[\epsilon_{\alpha}+\epsilon_{\beta}-2\mathrm{E_{s}}])^{2}$, $\psi_{elas}2 = \frac{1}{2}c_{12}[\mathrm{E_s}-\epsilon_{\alpha} + \mathrm{E_s}-\epsilon_{\beta} + \frac{c_{12}}{c_{11}}(\epsilon_{\alpha}+\epsilon_{\beta}-2\mathrm{E_{s}})]^2$} On the Al substrate, the voltage plateaus shift lower by $\sim 10\%$ and $\sim 4\%$, respectively, for the $\delta-\gamma$ and $\gamma-\omega$ stages, when compared to the reference voltage plateaus for a free standing $\mathrm{Li_xV_2O_5}$. Similarly, on the Au substrate, large voltage plateau shifts of $8\%$ and $3\%$ are observed, respectively, for the $\delta-\gamma$ and $\gamma-\omega$ stages. In these voltage shifts, we note that $\psi_{elas}1$ plays an important role and accounts for up to $\sim 65\%$ of the total voltage plateau shifts. We attribute this effect to the increase in misfit strains between the imposed film strain and the preferred lattice strains (e.g., ($\mathrm{E_s}-\epsilon_{\alpha}$), ($\mathrm{E_s}-\epsilon_{\beta}$)) at the $\delta-\gamma-\omega$ phase transformations.

It should also be noted that the final $\gamma-\omega$ phase transformation is accompanied by severe puckering and large structural changes that are known to amorphize the electrode \citep{christensen2018structural,JM1992Electrochemical}. This irreversible structural damage to the electrode can be prevented by modeling $\mathrm{V_2O_5}$ electrodes as thin-films with suitable film strains. Specifically, film stresses of $\sim2\mathrm{GPa}$ (i.e., $\mathrm{E_s\approx1\%}$) lowers the $\gamma-\omega$ voltage plateau to below $2$V and can thus suppress phase transformation when operated in the 2--4V voltage window. This strain engineering of thin-film $\mathrm{V_2O_5}$ electrodes facilitates its structural reversibility.

In short, these changes of voltage plateau caused by film strains can facilitate reversible phase transformation in two ways: First, the downward sloping voltage plateaus suggest suppressed phase separations. This suppression of phase separation reduces coherency stresses, which are known to nucleate microcracks in electrodes. Second, the shifting of voltage plateaus to lower values and outside of the typical operational window (2.5--4 V) prevents phase transformation. This shift in voltage plateaus can be engineered to prevent phase transformations with large volume changes. \textcolor{black}{We will apply these concepts to engineer reversible phase transformation in $\mathrm{V_2O_5}$ electrode in Section 3.}

\subsection{Stress response}
We assume a linear elastic cubic relation and describe the stress tensor $\boldsymbol{\sigma}$ as a function of the stiffness tensor $\mathbb{C}$, the preferred strain tensor $\mathbf{E}_\mathrm{0}$, and the Li-content $c$:
\begin{align}
\boldsymbol{\sigma}(c,\mathbf{E}) & = \frac{\partial\Psi}{\partial\mathbf{E}}  \nonumber\\
& =  \mathbb{C}:[\mathbf{E-E}_{0}(c)]
\label{eq:stress evolution}
\end{align}

The stress components are individually given in Eq. (14) of the Appendix. As discussed previously we assume a single crystal electrode and neglect the effect of grain structure and the kinetics of phase transformation on the stress response. We model the preferred strain tensor to describe the affine lattice parameter changes in $\mathrm{Li_xV_2O_5}$ during phase transformation as a function of Li-content $c$. We calibrate the preferred strains with the experimental lattice parameter measurements for $\mathrm{Li_xV_2O_5}$ as shown in Fig.~\ref{fig:strain-fit}(a-c). We fit the preferred strain data using a third-order polynomial that captures the general trend of preferred strains as a function of Li-content. While a higher-order polynomial would capture the abrupt changes in lattice strains (shown by the experimental data in Fig.~\ref{fig:Stress-Evolution}), we use a lower order polynomial to prevent overfitting of the strain data. Although this polynomial approximation of the stresses does not capture the abrupt changes in the in-plane stresses as measured experimentally, it provides a continuum formulation of in-plane stresses as a function of Li-content and crystal orientation.

\begin{figure}[H]
\begin{centering}
\includegraphics[width=\textwidth]{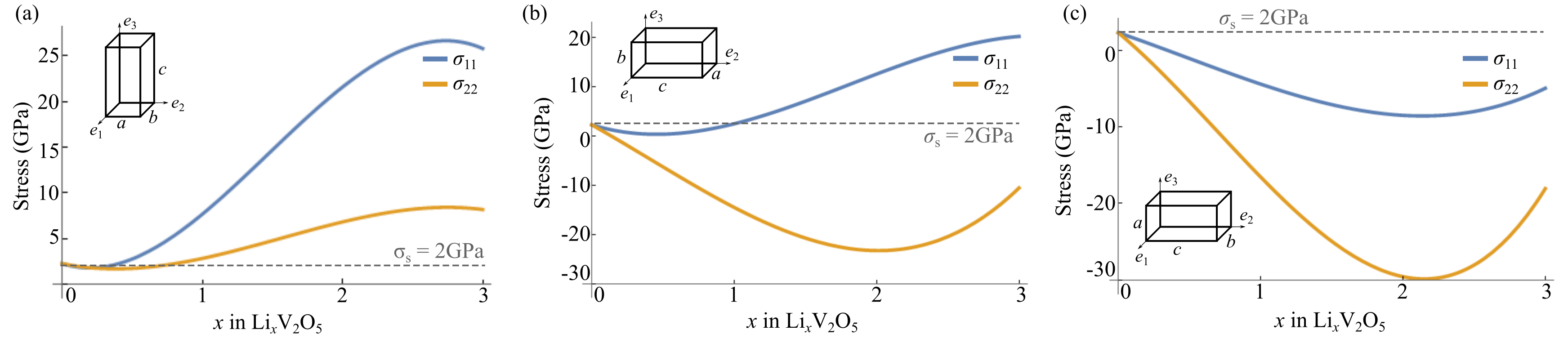}
\par\end{centering}
\caption{\label{fig:stress-theory} In-plane stresses $\sigma_{11}$ and $\sigma_{22}$ for a $\mathrm{Li_\mathit{x}V_2O_5}$ thin-film electrode as a function of Li-content for three crystal orientations: (a) with the $c$--axis perpendicular to the plane of the film, (b) with the $a$--axis perpendicular to the plane of the film, and (c) $b$--axis perpendicular to the plane of the film. The inset schematics illustrate the crystal orientations and the dashed line corresponds to the film stress $\sigma_\mathrm{s}$.}
\end{figure}

Fig.~\ref{fig:stress-theory}(a-c) depicts the in-plane stress components $\sigma_{11}, \sigma_{22}$ as a function of Li-content for three different crystallographic orientations of $\mathrm{Li_xV_2O_5}$. To be specific,  the crystal in Fig.~\ref{fig:stress-theory}(a) is oriented such that its $c-$axis is perpendicular to the thin-film plane. In this case, the in-plane stress $\sigma_{11}$ goes up with increasing Li-content. \textcolor{black}{By contrast, for a different crystal orientation as in Fig.~\ref{fig:stress-theory}(c), $\sigma_{11}$ decreases initially and then shows a upward tendency with increasing Li-content. A similar variation of $\sigma_{22}$ is also observed.} These contrasting stress responses for different in-plane stress components and for different crystal orientations will affect the net stress evolution in a polycrystalline material. In such cases, a detailed measurement of grain orientations in the electrode is necessary to estimate the net stress response.

\section{Experiment}
In this section, we present results from an experiment investigating the effect of film strains on the reversible cycling of $\mathrm{V_2O_5}$ thin-film electrodes. As such this study serves as an experimental test of the theoretical results presented in Section 2. 

\subsection{Methods}
We designed four custom-made electrochemical cells, each comprising of a thin-film $\mathrm{V_2O_5}$ cathode, Li-metal anode, and a liquid electrolyte.\footnote{equimolar mixture of Ethylene Carbonate, Dimethyl Carbonate, and 1M $\mathrm{LiPF_6}$.} In two of the cells, we deposited $\mathrm{V_2O_5}$ on Aluminum-coated-quartz (Al), and in the remaining two cells we deposited $\mathrm{V_2O_5}$ on Gold-coated-quartz (Au) substrate, see Fig.~\ref{fig:Schematic-illustration-of-thin-film}(a). \textcolor{black}{We next annealed the thin-film cathodes at 550$^\circ$C in the air for three hours. On cooling the cathodes the $\mathrm{V_2O_5}$ thin-film shrinks more than the quartz glass substrate, see Fig.~\ref{fig:Schematic-illustration-of-thin-film}(b-c). This difference in lattice geometries between the thin-film and the substrate gives rise to a lattice misfit and induces tensile strains in the $\mathrm{V_2O_5}$ electrodes. Using the MOSS technique, we measure the in-plane stress values of the $\mathrm{V_2O_5}$ film on Al and Au substrates to be 2GPa and 1GPa respectively. In this work, we do not study the mechanistic cause for the difference in in-plane stresses for the Al and Au substrates, however, we use this difference to investigate the effect of film stresses on reversible cycling of $\mathrm{V_2O_5}$ thin films.} The $\mathrm{V_2O_5}$ films on both Al and Au substrates are crystalline in the initial state and were characterized using transmission electron microscopy (TEM), scanning electron microscopy (SEM) and X-ray diffractometer (XRD), see Fig.~\ref{fig:potentiostatic-cycling}(a). For further details on the thin-film preparation, characterization, and MOSS technique, please see Appendix \ref{Thin-film experiment}. 

Next, we divided the cells into two groups (each group containing two cells with a $\mathrm{V_2O_5}$ cathode deposited on Al and Au substrates), and tested our theory that film strain enhances the structural reversibility of thin-film cathodes during electrochemical cycling. In the first group, we potentiostatically cycled the cells at constant voltages 4V, 3V, and 2V. Each of these holds were maintained until the films were equilibrated. After three charge/discharge cycles we conducted XRD measurements and SEM imaging for the  $\mathrm{V_2O_5}$ films on Al and Au substrates, respectively. In the second group, we galvanostatically cycled the cells at a C-rate of C/20 and typically in the voltage window of 2-4V. During galvanostatic cycling, we measured the voltage and stress values for the $\mathrm{V_2O_5}$ films on Al and Au substrates, respectively. Further details on the electrochemical cycling and the in-situ voltage-stress measurement techniques are described in Appendix \ref{Thin-film experiment}.

\subsection{Potentiostatic cycling}

\begin{figure}[H]
\begin{centering}
\includegraphics[width=\textwidth]{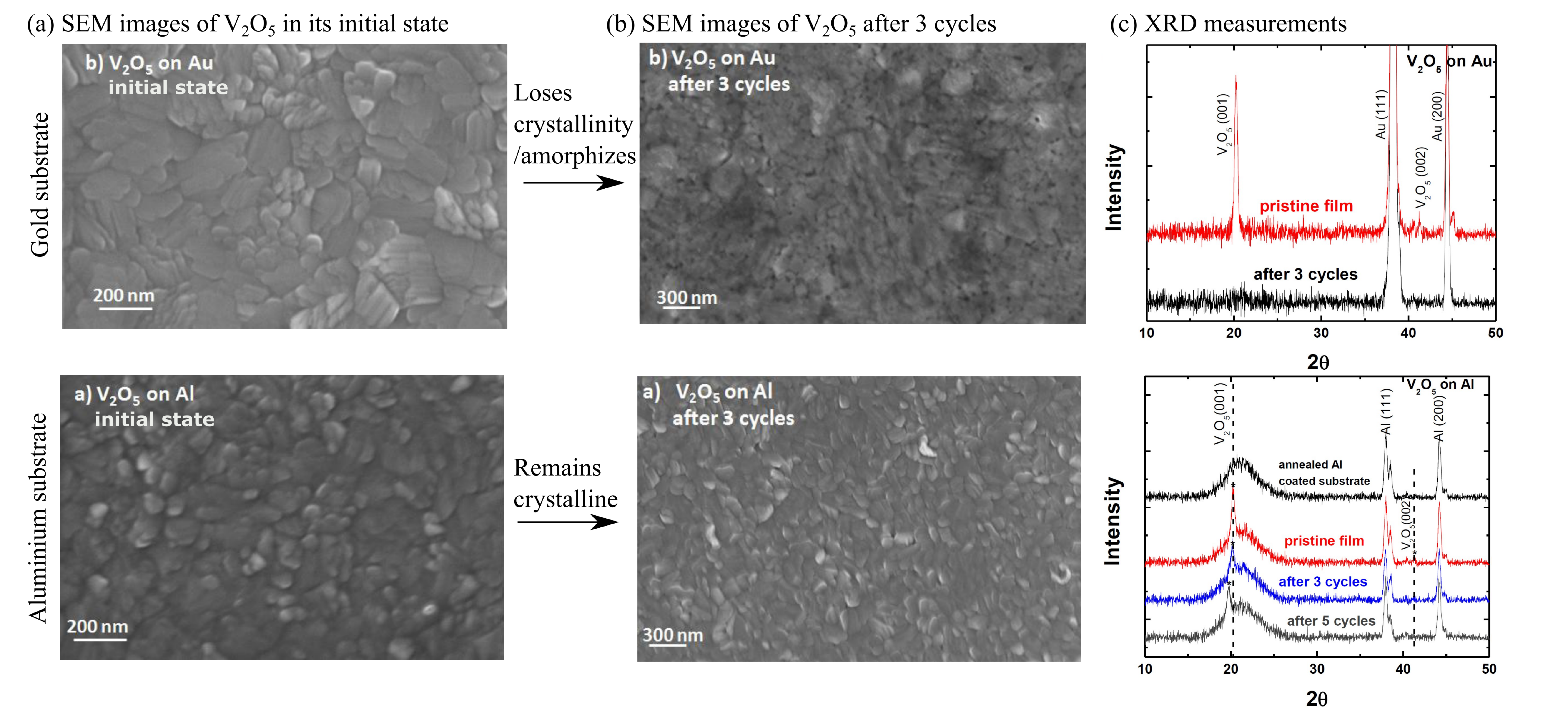}
\par\end{centering}
\caption{\label{fig:potentiostatic-cycling}(a) SEM images of the $\mathrm{V_{2}O_{5}}$ electrode surface deposited on a Au substrate (top) and an Al substrate (below). These images show the microstructure on $\mathrm{V_{2}O_{5}}$ surface before electrochemical cycling. (b) SEM images of $\mathrm{V_{2}O_{5}}$ surface after potentiostatic cycling (3 cycles). Our experiments show that the $\mathrm{V_{2}O_{5}}$ film deposited on Al retains crystallinity on cycling while the $\mathrm{V_{2}O_{5}}$ film deposited on Au loses crystallinity or amorphizes on potentiostatic cycling. (c) The X-ray diffraction measurements of the $\mathrm{V_{2}O_{5}}$ film on Al and Au substrates cycled potentiostatically. In the Al case, the X-ray peak corresponding to $\mathrm{V_{2}O_{5}}$ (001) broadens upon repeated cycling. But in the Au case, the X-ray peak disappears after 3 cycles. These measurements further confirm the loss of crystallinity in $\mathrm{V_{2}O_{5}}$ film deposited on Au.}
\end{figure}

Fig.~\ref{fig:potentiostatic-cycling}(a-c) shows the SEM images and XRD measurements of the $\mathrm{V_2O_5}$ cathode on Au and Al substrates respectively. The SEM images in Fig.~\ref{fig:potentiostatic-cycling}(a-b) show the $\mathrm{V_2O_5}$ electrode surface before and after three potentionstatic cycles. The $\mathrm{V_2O_5}$ electrode on Au ($\sim1\mathrm{GPa}$ film stress) amorphizes after three cycles while the $\mathrm{V_2O_5}$ electrode on Al ($\sim2\mathrm{GPa}$ film stress) remains crystalline after three cycles. The XRD measurements in Fig.~\ref{fig:potentiostatic-cycling}(c) show the structural evolution of crystallographic peaks corresponding to $\mathrm{V_2O_5}$ cathode on Au and Al substrates respectively. Specifically, the $\mathrm{V_2O_5}$ cathode on Au substrate loses crystallinity after three cycles. This is confirmed by the loss of the (001) peak in its XRD measurement, see Fig.~\ref{fig:potentiostatic-cycling}(c). By contrast, for the $\mathrm{V_2O_5}$ cathode on Al substrate, the (001) peak gradually broadens and shifts towards lower theta values with repeated cycling. But this cathode still remains crystalline even after five cycles and maintains its orthorhombic structure. Overall, these potentiostatic experiments show that the increased film straining enhances the structural reversibility of $\mathrm{V_2O_5}$ electrodes.

\subsection{Galvanostatic cycling}

\begin{figure}[H]
\begin{centering}
\includegraphics[width=\textwidth]{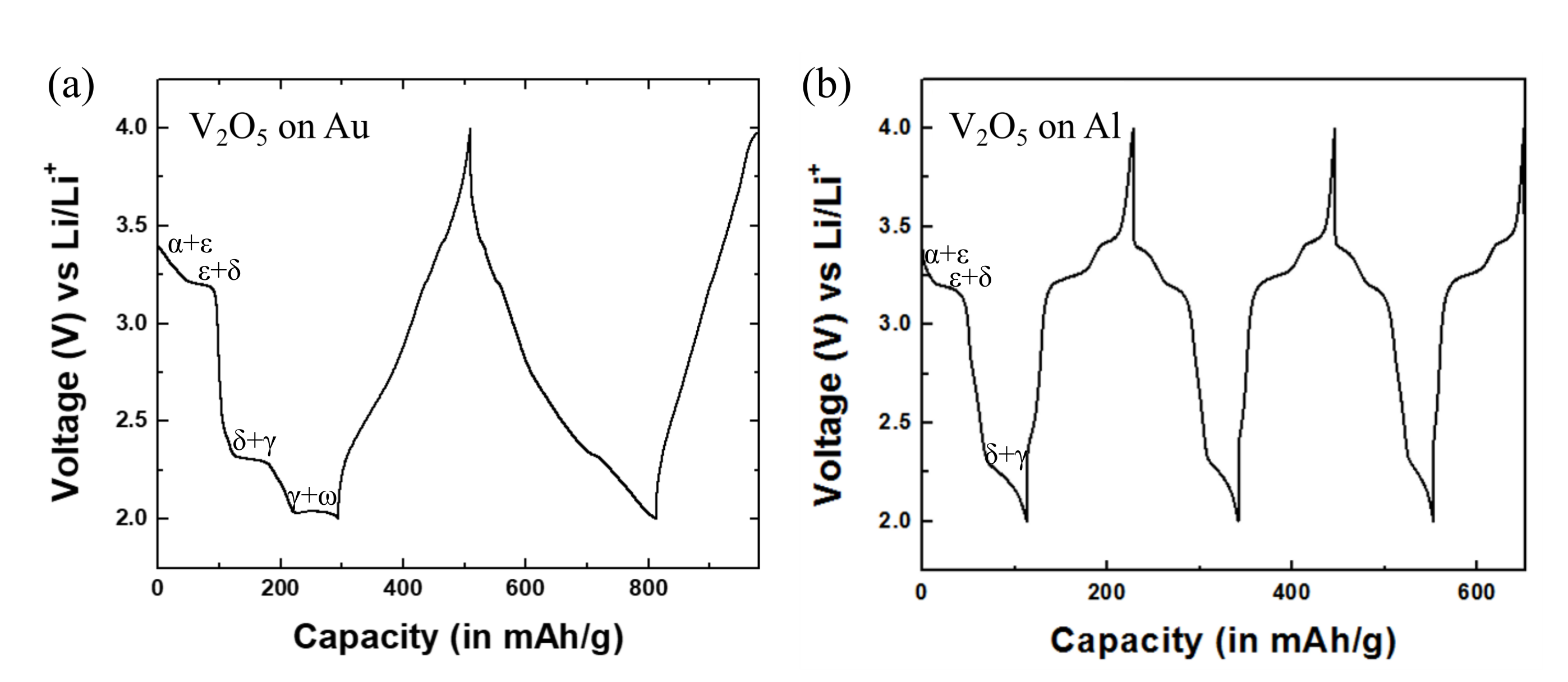}
\par\end{centering}
\caption{\textcolor{black}{Voltage-capacity plots of $\mathrm{V_2O_5}$ cathode deposited on Au and Al substrates and cycled galvanostatictically in a fixed 2--4V voltage window.  Both the thin-film cathodes were cycled with a constant current density of $\mathrm{0.2\mu A/cm^{2}}$ and at a rate of C/20. (a) The voltage curve for $\mathrm{V_2O_5}$ on Au substrate with $\mathrm{\sigma_s \approx 1 GPa}$ shows four voltage plateaus in the first cycle. These voltage plateaus are not observed in subsequent cycles showing structural irreversibility. (b) The voltage curve for $\mathrm{V_2O_5}$ on Al substrate with $\mathrm{\sigma_s \approx 2 GPa}$ shows only three voltage plateaus in the first cycle. These voltage plateaus are observed in the subsequent cycles showing structural reversibility.}}
\label{fig:Galvanostatic-Cycling}
\end{figure}

Fig.~\ref{fig:Galvanostatic-Cycling}(a-b) shows the in-situ voltage measurements of $\mathrm{V_2O_5}$ on Au and Al substrates respectively. Both the electrochemical cells were cycled galvanostatically, under a constant current density of $\mathrm{0.2\mu A/cm^{2}}$ and at a rate of C/20, in a fixed voltage window of 2--4V. The voltage traces in this range demonstrate that altering the initial stress state leads to substantial differences in the electrochemical response. This comparison also provides an explanation for the amorphization observed during potentiostatic cycling with an Au current collector in Fig.~\ref{fig:potentiostatic-cycling}. Here the voltage responses show plateaus corresponding to the phase transformations at different Li-intercalation stages. In Fig.~\ref{fig:Galvanostatic-Cycling}(a), the $\mathrm{V_2O_5}$ cathode on Au substrate undergoes four distinct phase transformation stages that are characterized by one voltage slope (at 3.4V) and three voltage plateaus (at 3.20V, 2.31V, and 2.04V). The final $\gamma-\omega$ phase transformation stage is accompanied by a large volume change that amorphizes the $\mathrm{V_2O_5}$ cathode. This amorphization contributes to the structural irreversibility of the cathode and thus no distinct voltage plateaus are observed in the subsequent discharge cycles. These features of structural irreversibility and loss of voltage plateaus on the $\gamma-\omega$ phase transformation in $\mathrm{V_2O_5}$ are consistent with results on $\mathrm{V_2O_5}$ amorphization reported by other researchers \citep{whittingham1976role,christensen2018structural}.

By comparison, with a larger tensile film stress ($\mathrm{\sigma_s \approx 2GPa}$), the $\mathrm{V_2O_5}$ cathode on the Al substrate undergoes only three phase transformation stages that are characterized by voltage slopes/plateaus at 3.33V, 3.18V, and 2.12V respectively. These phase transformations occur at lower voltage values when compared to their corresponding transformations on the Au substrate. The final $\gamma-\omega$ phase transformation stage is not observed in $\mathrm{V_2O_5}$ on Al substrates in the 2--4V operational window. Consequently, the large volume changes associated with the $\gamma-\omega$ phase transformation does not occur in the 2--4V voltage window, and the structural reversibility of the cathode is preserved. Thus, in the subsequent discharge cycles on Al substrate, the $\mathrm{V_2O_5}$ electrode retains its structural reversibility as shown by the recurrence of the voltage plateaus in Fig.~\ref{fig:Galvanostatic-Cycling}(b).

\subsection{Voltage plateaus and in-plane stresses}
\begin{figure}[H]
\begin{centering}
\includegraphics[width=1\textwidth]{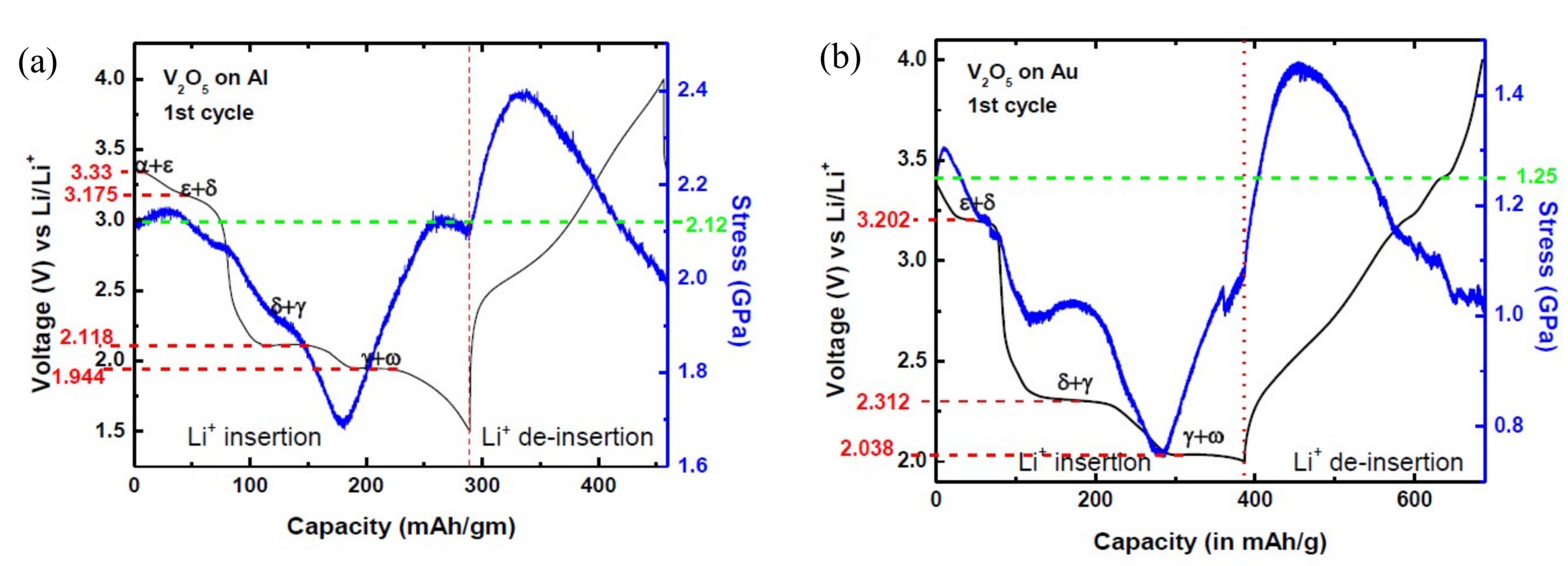}
\par\end{centering}
\caption{In-situ voltage and stress measurements for the crystalline $\mathrm{V_{2}O_{5}}$
film on (a) Al and (b) Au substrates respectively. Both $\mathrm{V_{2}O_{5}}$ films are cycled galvanostatically, with a constant current density of $\mathrm{0.2\mu A/cm^{2}}$ and at a rate of C/20. The voltage plateaus corresponding to multiple phase transformations in $\mathrm{V_{2}O_{5}}$, and the inplane stresses measured using the MOSS technique are shown in subfigures (a) and (b). 
\label{fig:Stress-Evolution}}
\end{figure}

We next do a deep discharge of the $\mathrm{V_2O_5}$ cathodes on both the Au and Al substrate until all four phase transformations have occurred. \textcolor{black}{That is, unlike in Section~3.3, we do not fix the operational voltage window to 2--4V, but do a deep discharge until all four phase transformations occur in $\mathrm{V_2O_5}$ on both the Au and Al substrates.} During these discharge cycles we measure both the voltage and stress responses as a function of Li-content, see Fig.~\ref{fig:Stress-Evolution}(a-b).

\textcolor{black}{\uline{Voltage Plateaus}: Fig.~\ref{fig:Stress-Evolution}(a-b) shows four distinct voltage plateaus corresponding to the $\alpha-\epsilon-\delta-\gamma-\omega$ phase transformations in $\mathrm{V_2O_5}$ cathodes on both the Au and Al substrates.\footnote{Note that the $\mathrm{Li_xV_2O_5}$ cathode expands in volume (i.e., positive volume change) during initial intercalation up to $\mathrm{x}\leq2$, and then contracts in volume (i.e., negative volume change) during the final stages of intercalation for $\mathrm{x}>2$, see Fig.~\ref{fig:strain-fit}(a-c) in the Appendix.} Here, we note two distinguishing features of the voltage curves: The first feature is that the voltage plateaus for  $\mathrm{V_2O_5}$ on the Al substrate are lower than the corresponding voltage plateaus on the Au substrate in Fig.~\ref{fig:Stress-Evolution}(a). Take the $\gamma-\omega$ stage as an example, the voltage plateau on the Al substrate is at 1.94V which is $\sim5\%$ smaller than the voltage plateau at 2.038V on the Au substrate. Similarly, the voltage plateau at the $\delta-\gamma$ phase transformation stage is lowered to 2.118V on the Al substrate.\footnote{\textcolor{black}{Note, that the $\delta-\gamma$ voltage plateau of $\mathrm{Li_xV_2O_5}$ on the Au substrate shifts by only $\sim 3\%$ compared with 2.374V of the free-standing one. We explain this surprisingly small voltage shift to the combined tensile and compressive stress responses measured during the $\delta-\gamma$ phase transformation stage, see Fig.~\ref{fig:Stress-Evolution}(b).}} These voltage shifts in Fig.~\ref{fig:Stress-Evolution}(a-b) are consistent with our theoretical predictions in Section~2.2. We explain the downward (and relatively large) shifts of the voltage plateaus to the larger misfit stresses on the Al substrate. These large stresses combined with the negative volume change (e.g., for $\gamma-\omega$ phase transformation) contribute to the downward shifts of the voltage plateaus (from Eq.~\ref{eq:ChemicalPotential}).}

\textcolor{black}{The second feature is that the voltage plateau shifts downward for both the $\delta-\gamma$ and $\gamma-\omega$ phase transformation stages despite their opposing stress responses (and/or volume changes) during intercalation. That is, the $\delta-\gamma$ voltage plateau at 2.31V on Au substrate is lowered to 2.12V on the Al substrate, despite the positive volume change during the $\delta-\gamma$ phase transformation. We attribute this discrepancy to the polycrystalline nature of the $\mathrm{V_2O_5}$ electrode in our experiment (i.e., electrode contains grains with multiple orientations), and the large tensile film stresses that contribute to the net lowering of the voltage plateaus. For example, the film stress in $\mathrm{V_2O_5}$ on Al substrate ($\mathrm{\sigma_s \approx 2 GPa}$) is nearly twice as large as the film stress in $\mathrm{V_2O_5}$ on the Au substrate ($\mathrm{\sigma_s \approx 1 GPa}$). This difference in in-plane stresses lowers the voltage values for phase transformations despite their positive volume change as shown theoretically in Section~2.2 and 2.3.}

\textcolor{black}{\uline{Stress response}: During Li-insertion, the stress responses measured using the MOSS technique exhibit features that are similar to both the Al and Au current collectors, see Fig.~\ref{fig:Stress-Evolution}(a-b). For example, the cathodes in both cases show compressive stresses during the $\alpha-\epsilon-\delta-\gamma$ phase transformations $(\mathrm{x}<2)$, which suggest positive volume change (expansion) of the cathode; During the $\gamma-\omega$ phase transformation stage $\mathrm{x>2}$ we observe tensile stresses that suggests a negative volume change (contraction) of the cathode. Surprisingly, during Li-extraction the stress response is tensile during the initial stages. For reversible elastic behavior in the $\gamma-\omega$ phase transformation, one would expect the Li-extraction to be compressive. With continual delithiation $(x<1.8)$, the stresses eventually decrease towards their initial states.}

\textcolor{black}{We next note that the net stress response, measured using the MOSS technique in Fig.~\ref{fig:Stress-Evolution}, is an order of magnitude smaller than our theoretical predictions for an ideal single-crystal thin-film electrode, see Fig.~\ref{fig:stress-theory}(a-c). We attribute this difference in stress responses between theory and experiment to the polycrystalline texture of the electrode. As discussed in Section~2.4, single crystals with different crystallographic orientations generate distinct in-plane stresses. In our experiment, the polycrystalline electrode contains grains of different orientations and we measure its net stress response using the MOSS technique. The average stress response in Fig.~\ref{fig:Stress-Evolution}(a-b) first decreases during lithiation (for $x \leq 2$) and increases thereafter (for $x > 2$). This stress evolution as a function of Li-content is similar to the stresses predicted by our analytical model for a single crystal $\mathrm{Li_xV_2O_5}$ with $c-$axis in the $\mathrm{e_1-e_2}$ plane, see Fig.~\ref{fig:stress-theory}(b-c). This similarity in stress responses suggests that the majority of the grains in the polycrystalline electrode are oriented with their $c-$axis in the $\mathrm{e_1-e_2}$ plane. In a future study, we plan to map the stress evolution as a function of the electrode's individual grain orientations.}

\vspace{5mm}
Overall, we note three key observations in our experiments: First, in the potentiostatic experiment, the $\mathrm{V_{2}O_{5}}$ electrode on Al substrate retains crystallinity upon repeated cycling in the 2--4V operational window while the $\mathrm{V_{2}O_{5}}$ electrode on Au substrate loses crystallinity (or amorphizes) after three cycles. Second, in the galvanostatic experiment, the voltage plateaus corresponding to $\mathrm{V_{2}O_{5}}$ phase transformations are shifted to lower values with an increase in the initial film strain. This shift of voltage plateaus, specifically the lowering of $\gamma-\omega$ plateau, expands the operational voltage window for lithium batteries. Third, the in-plane stresses, in both potentiostatic and galvanostatic experiments, evolve as a function of Li-content and are anisotropic in the $\mathrm{Li_\mathit{x}V_2O_5}$ thin-film electrode.  

\section{Discussion}
Our theoretical and experimental studies demonstrate that the film strain enhances the reversible cycling of thin-film $\mathrm{V_2O_5}$ cathodes in the 2--4V voltage window. The theoretical calculations predict that the film strains shift the voltage plateaus to lower values and circumvent the enormous volume changing phase transformation ($\mathrm{Li_2V_2O_5-Li_3V_2O_5}$). This prediction was confirmed in our experiments that demonstrate structural reversibility of $\mathrm{V_2O_5}$, with film stresses $\sim2\mathrm{GPa}$ and across a wider voltage window 2--4V. In the remainder of this discussion, we discuss the limiting conditions on our theoretical and experimental framework and then compare our results with prior work on phase transformation electrodes.  

Two features of this work limit the conclusion we can draw about the effect of film strain on reversible cycling of thin-film electrodes. First, we derive the theoretical model by assuming the thin-film electrode to be a single crystal, and thus our predictions on voltage curves and stress responses are subject to the shortcomings associated with the presence of grain boundaries and other defects commonly found in a polycrystalline material. Second, we account for the effect of film stress in the variational derivative of the free energy (i.e., chemical potential) and do not consider the effect of diffusion kinetics on voltage measurements. Prior research shows that the internal stress states (e.g., arising from film strains) can alter diffusion rates \citep{ganser2019extended,sultanova2021microscale} and affect the mechano-chemistry of batteries \citep{bucci2017effect}. In order to capture this effect, the diffusion kinetics would need to be coupled with the evolving stresses in the electrode. This effect is currently not captured in the present theory and we plan to address it in a future study.

Our findings differ from prior research on suppressing phase transformation in intercalation electrodes. First, unlike previous studies \citep{shimizu1993electrochemical,
baddour2012structural}, we do not constrain the operational voltage window between 3--4V to prevent the $\gamma-\omega$ phase transformation in $\mathrm{V_2O_5}$ electrodes. The film strains enable us to electrochemically cycle$\mathrm{V_2O_5}$  across a wider voltage window 2--4V, which increases the specific capacity of the battery. Prior research on thin-film $\mathrm{V_2O_5}$ electrodes similarly report a wider voltage window \citep{mcgraw1998next}. Second, we need not engineer electrode geometry (for example nano-sized particles) or Li-kinetics (increased C-rates) to suppress phase separation. The film strains can be engineered to shift the voltage plateaus to lower values, which enables the reversible cycling of $\mathrm{V_2O_5}$ despite its thin-film geometry or small C-rates.

Compared to other electrode materials such as Si-Li alloys, our findings show that, for $\mathrm{V_2O_5}$, the elastic energies have a more complex impact on the two-phase equilibria that are associated with voltage plateaus. In Si-Li alloys deleterious phase transformations are known to limit their voltage range for reversible cycling. That is, voltages below 0.05 V are generally avoided to prevent its transformation to the crystalline $\mathrm{Li_4Si_{15}}$ phase \citep{obrovac2004structural}. Since this transformation involves a structural change at a fixed composition, the voltage plateaus that feature prominently in $\mathrm{V_2O_5}$ are not observed. However, elastic energy effects are still potentially relevant for Si-Li alloys and tensile strains should inhibit the formation of the $\mathrm{Li_4Si_{15}}$ phase which is accompanied by a volume decrease. Another distinguishing feature between $\mathrm{V_2O_5}$ and Si-Li alloys is that the stress-induced voltage shifts in the $\mathrm{V_2O_5}$ experiments are relatively large when compared to Si-Li alloys. For example, in a single phase electrode material like Si, a 0.5 GPa change in the in-plane stress corresponds to a voltage shift of roughly 0.03 V, whereas in our experiments the comparable shifts for the $\delta-\gamma$ and $\gamma-\omega$ plateaus are much larger than this. These examples show that elastic energies play an important role in regulating phase transformation voltages in electrodes beyond $\mathrm{V_2O_5}$.

More generally, the present work establishes a theoretical and experimental framework to enhance the lifespans of intercalation electrodes. Specifically, we show that film strains can modulate the free energy landscape of phase transformation electrodes. These strains can be engineered precisely (e.g., by using shape-memory alloy substrates \citep{muralidharan2017tunable}) to control the voltage plateaus for phase transformation, and can be calibrated to offer wider voltage windows for battery operation.  Beyond $\mathrm{V_2O_5}$, the study serves as a theoretical guide to tailor film strains on thin-film electrodes and could be used as a design tool to engineer the mechano-chemistry of batteries. Applying this to thin film electrodes is directly relevant in certain microbatteries.  Strain engineering strategies are also potentially relevant in a broader range of practical electrodes, and here implementation will require approaches that reflect the structures of the specific materials.

\section{Conclusion}
{To conclude, we use a combination of theory and experiments to show that film strains regulate phase transformations in thin-film intercalation electrodes and that film strains can be engineered to prevent chemo-mechanical degradation of electrodes. Specifically, our findings show that tensile strains lower the voltage for phase transformation in $\mathrm{V_2O_5}$ thin-film electrodes and facilitate their reversible cycling across a 2--4V voltage window that is wider than previously achieved. More broadly, our research shows a novel route of engineering film strains of thin-film electrodes that make intercalation electrodes with longer lifespans possible.

\section{Acknowledgement}
  The authors acknowledge the High-Performance computing center at the University of Southern California for providing resources that contributed to the research results reported within this paper. A.R.B acknowledges the support of the Provost Fellowship and start-up funds from the University of Southern California. Research at Brown University was supported by the Ceramics Program at the National Science Foundation (DMR-1832829).} 

\newpage
\section{Appendix}

\subsection{Theoretical calculations}\label{Theoretical calculations}

\subsubsection*{Phase transformation in $\mathrm{Li_xV_2O_5}$}

\begin{figure}[H]
    \centering
    \includegraphics[width=0.8\textwidth]{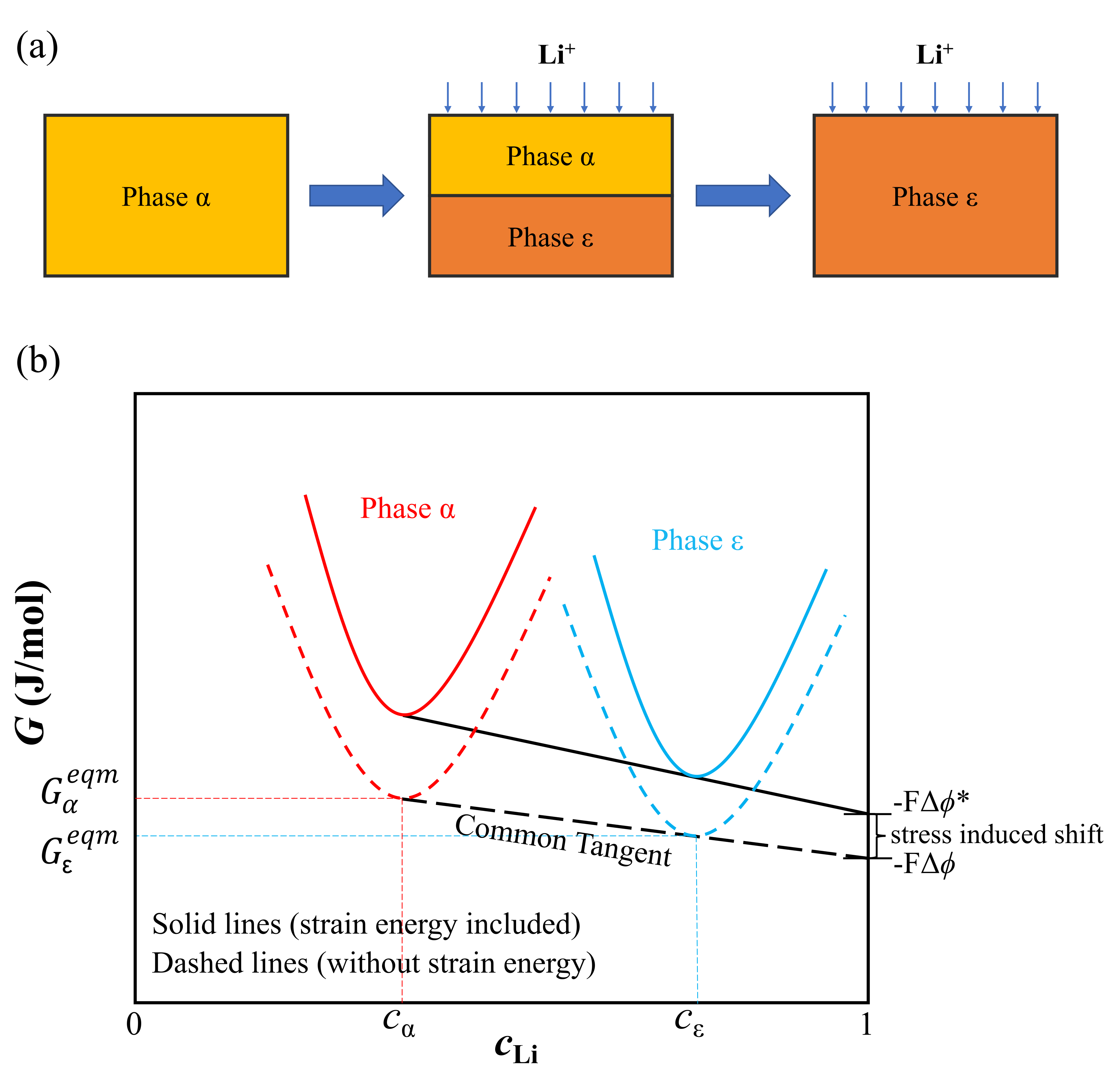}
    \caption{Schematic showing (a) the representative phase transformation (from phase $\alpha$ to phase $\epsilon$) upon lithiation in a intercalation material, and (b) graphical representation of the free energy curves of two phases, $\alpha$ and $\epsilon$ with (solid lines) and without (dashed lines) strain energy contributions. $G_{\alpha}^{eqm}$ and $G_{\epsilon}^{eqm}$ are equilibrium molar Gibbs free energy for phase $\alpha$ and $\epsilon$ in the case without strain energy contribution. $c_\alpha$ and $c_\epsilon$ are the equilibrium Li-content in the $\alpha$ and $\epsilon$ phases respectively. $\mathrm{F}\Delta \phi^*$ and $\mathrm{F}\Delta \phi$ are the chemical driving forces with and without the strain energy, respectively.}
    \label{fig:molar Gibbs free energy}
\end{figure}

Crystalline $\mathrm{V_2O_5}$ undergoes multiple phase transformations as a function of Li-content that are characterized by constant voltage plateaus \citep{christensen2018structural}. Each voltage plateau corresponds to the co-existence of two phases of $\mathrm{Li_xV_2O_5}$ as shown in Fig. \ref{fig:Galvanostatic-Cycling}. For example, consider the $\alpha-\epsilon$ phase transformation stage (see Fig.~\ref{fig:molar Gibbs free energy}(a)), at the beginning of this voltage plateau Li is inserted into the host material that induces phase transformation from the $\alpha$ to the $\epsilon$ phase. On further lithiation the composition of the $\alpha$ or $\epsilon$ phases does not change, however the $\epsilon$ phase grows in size at the expense of the $\alpha$ phase. This growth is marked by the movement of a phase boundary and both the $\alpha,\epsilon$ phases coexist on this voltage plateau. At the end of the voltage plateau all of the $\alpha$ phase is transformed to the $\epsilon$ phase. This two phase behaviour is widely reported for the $\mathrm{Li_xV_2O_5}$ compound for all its phase transformation stages \citep{wang2006nanostructured,christensen2018structural}.

In our analytical calculations we assume a double-well potential to describe the two phase coexistence at each voltage plateau in $\mathrm{Li_xV_2O_5}$. Other possible configurations, such as a pre-existing nucleus of a third phase at a grain boundary are possible, however, in the absence of specific data on these microstructures, the two phase configuration described above provides a fundamental framework to investigate the role of elastic energy on the voltage plateaus. Similar two phase configurations are assumed to model phase transformations in other materials, such as ferroelectrics, shape memory alloys, and ferromagnets that show series of solid-solid phase transformation stages as a function of temperature \citep{chen2008phase,mamivand2013review}. In line with previous research, we model the two-phase coexistence of $\mathrm{Li_xV_2O_5}$ at each voltage plateau using a double-well potential.

We next describe how the solubility limits and enthalpy of mixing are derived for a two phase transformation in the $\mathrm{Li_xV_2O_5}$ system. These thermodynamic properties of the material are fitted with the voltage measurements for $\mathrm{Li_xV_2O_5}$ at each transformation stage. Fig. \ref{fig:molar Gibbs free energy}(b) shows a free energy curves for a representative two phase transformation. In our analysis we describe this free energy curve using the standard regular solution model, see Eq. (6). The regular solution model has been previously used to describe phase transformations in intercalation electrodes, such as LiFePO$_4$, LiCoO$_2$, $\mathrm{Li_xV_2O_5}$ \citep{tang2011anisotropic,nadkarni2019modeling,bai2021chemo}. We calibrate the coefficients of the regular solution model to fit with the experimentally measured thermodynamic and elastic properties of the material.

For example, in Fig.~\ref{fig:molar Gibbs free energy}(b), the $c_\alpha$ and $c_\epsilon$ are the equilibrium Li-content in the $\alpha$ and $\epsilon$ phases respectively.\footnote{In our calculation we normalize this Li-content to vary as $0 \leq c \leq 1$.} These equilibrium Li-contents are derived from the voltage curves and serve as the solubility limits for the binary model. In our calculations we normalize the solubility limits for each two-phase transformation stage, i.e., $0 \leq c \leq 1$. Next, the chemical potential difference between the active material (i.e., $\mathrm{V_2O_5}$) and the counter electrode (i.e., Li-metal in our experiments) serves as the chemical driving force for Li-ion exchange in the active material. We model this driving force in terms of the Gibbs free energy change per mole of reaction: $\Delta \psi = -z\mathrm{F}\Delta \phi$. Here, $z = +1$ is the charge number of mobile Li-ion species, $\mathrm{F}$ is the Faraday constant, and $\Delta\phi$ is the potential difference between the two electrodes. The balance between the chemical and electrical forces on lithium exchange imply that the $\Delta \phi$ corresponds to the two phase equilibrium voltage plateaus in $\mathrm{Li_xV_2O_5}$ during phase transformation. That is the common tangent in Fig.~\ref{fig:molar Gibbs free energy}(b) corresponds to the equilibrium voltage for the two phase existence and is used to fit the values for the enthalpy of mixing. We use this strategy to fit the solubility limits and equilibrium voltage values for the remainder of the phase transformation stages (e.g., $\epsilon-\delta, \delta-\gamma, \gamma - \omega$ stages). While the exact shape of the curves and the energy barrier is approximate, this two-phase configuration provides the basic analytical framework to investigate the role of elastic energy on phase transformation voltages. Furthermore, a detailed model following the multi phase field approach \citep{steinbach2006multi} would be necessary to construct a free energy functional to describe a functional for a multi-phase transformation, however, this approach is beyond the current scope of our work.

\subsubsection*{Model parameters}
$\mathrm{V_2O_5}$ undergoes multiple phase transformation stages, and we assume piecewise functions to describe the energy landscape at each phase transformation stage. The thermodynamic and elastic constants are fitted to the electrochemical \citep{luo2018roadblocks} and lattice parameter measurements \citep{baddour2012structural,murphy1979lithium,cava1986structure,enjalbert1986refinement,cocciantelli1991crystal,dickens1979phase,meulenkamp1999situ,satto1999delta} accordingly. The table below lists the constants and polynomial functions used in our calculations.

\begin{table}[h]
\begin{centering}
\begin{tabular}{ll}
\hline 
\textbf{Thermodynamic constants} & \textbf{Value}\tabularnewline
\hline 
Gradient energy constant & $\kappa=3.084\times10^{-18}\mathrm{J/m}$\tabularnewline
Enthalpy of mixing & \tabularnewline
$\alpha-\epsilon$ & $\Omega_{1}=\textcolor{black}{5800}\mathrm{J/mol}$\tabularnewline
$\epsilon-\delta$ & $\Omega_{2}=\textcolor{black}{3571}\mathrm{J/mol}$\tabularnewline
$\delta-\gamma$ & $\Omega_{3}=\textcolor{black}{2890}\mathrm{J/mol}$\tabularnewline
$\gamma-\omega$ & $\Omega_{4}=\textcolor{black}{3587}\mathrm{J/mol}$\tabularnewline
Temperature & $T=298\mathrm{K}$\tabularnewline
\textcolor{black}{Gas constant}& $R=8.314\mathrm{J/mol\cdot K}$\tabularnewline
\hline 
 & \tabularnewline
\hline 
\textbf{Elastic constants} & \textbf{Value}\tabularnewline
\hline 
Stiffness constants & $c_{11}=\textcolor{black}{153.685}\mathrm{GPa}$\tabularnewline
 & $c_{12}=\textcolor{black}{65.865}\mathrm{GPa}$\tabularnewline
 & $c_{44}=\textcolor{black}{43.67}\mathrm{GPa}$\tabularnewline
\textcolor{black}{Preferred strain, see Fig.~\ref{fig:strain-fit}} & $\epsilon_{\alpha}=\textcolor{black}{0.024c^3-0.105c^2+0.035c}$\tabularnewline
 & $\epsilon_{\beta}=\textcolor{black}{0.001c^3-0.007c^2+0.015c}$\tabularnewline
 & $\epsilon_{\gamma}=\textcolor{black}{-0.025c^3+0.053c^2+0.119c}$\tabularnewline
\hline 
\end{tabular}
\par\end{centering}
\caption{\label{table:coefficients}Specific values of coefficients used for analytical calculations in section 2. These values correspond to phase transformations in $\mathrm{V_2O_5}$ electrode and are fitted to the experimental data provided by \cite{luo2018roadblocks}}.
\end{table}

\begin{figure}[H]
\begin{centering}
\includegraphics[width=\textwidth]{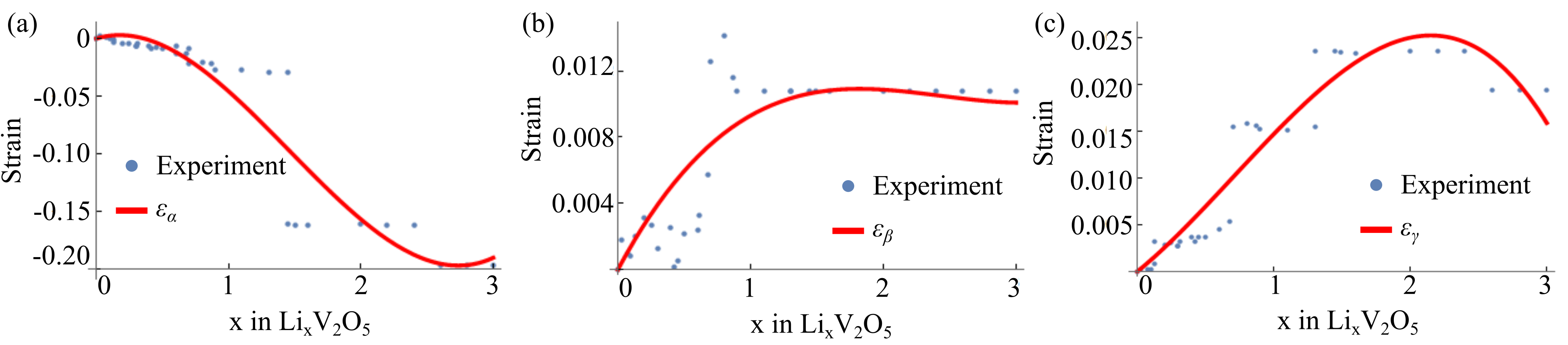}
\par\end{centering}
\caption{Lattice strains in $\mathrm{Li_\mathit{x}V_2O_5}$ as a function of Li-content. We fit polynomials with the strains provided by \cite{baddour2012structural,murphy1979lithium,cava1986structure,enjalbert1986refinement,cocciantelli1991crystal,dickens1979phase,meulenkamp1999situ,satto1999delta}.
\label{fig:strain-fit}}
\end{figure}

\subsection*{Analytical calculation of elastic energy with thin film conditions}
The general form of elastic energy $\psi_{elas}$:
\begin{align}
\psi_{elas}(c,\mathbf{E})&=\frac{1}{2}[{\bf E} - {\bf E}_{\rm 0}(c)] \ : \ \mathbb C [{\bf E} - {\bf E}_{\rm 0}(c)] \nonumber\\
&= \mathrm{C_{44}}[{\bf E} - {\bf E}_{\rm 0}({c})] \ : \  [{\bf E} - {\bf E}_{\rm 0}({c})] + {\rm{\frac{\mathrm{C_{12}}}{2}}}[({\bf E}-{\bf E}_{\rm 0}({c})]\textcolor{black}{\ : \ }{\rm tr}[{\bf E} - {\bf E}_{\rm 0}({c})]\bfI  \nonumber \\
& + \big(\frac{\mathrm{C_{11}-C_{12}}}{2}-\mathrm{C_{44}}\big) \sum_{i=1}^3 ({\bf e}_i \cdot [{\bf E} - {\bf E}_{\rm 0}({c})]{\bf e}_{\it i} )^2
\label{elastic_energy}
\end{align}

Note, $[{\bf E - \bf E}_{\rm 0}(c)] \textcolor{black}{\ : \ } [{\bf E - \bf E}_{\rm 0}(c)]$ expands to:
\begin{align}
[{\bf E} - {\bf E}_{\rm 0}(c)] \textcolor{black}{\ : \ } [{\bf E} - {\bf E}_{\rm 0}(c)]=\sum_{{\rm i=1}}^{\rm 3} ({\bfe}_{\it i} \cdot [{\bf E} - {\bf E}_{\rm 0}(c)]{\bfe}_{\it i} )^{\rm 2}) +\ 2(\mathrm{E^2_{12}} + \mathrm{E^2_{23}} + \mathrm{E^2_{13}})
\label{dotproduct}
\end{align}

By substituting Eq. \ref{dotproduct} in Eq. \ref{elastic_energy} we get:
\begin{align}
\psi_{elas}(c,\mathbf{E})&=\frac{1}{2}[{\bf E} -  {\bf E}_{\rm 0}(c)] \textcolor{black}{\ : \ } \mathbb C [{\bf E} - {\bf E}_{\rm 0}(c)]  \nonumber \\
& =2\mathrm{C_{44}}(\mathrm{E^2_{12}} + \mathrm{E^2_{23}} + \mathrm{E^2_{13}}) + \frac{\mathrm{C_{12}}}{2}[\mathrm{E_{11}} - \epsilon_{\alpha} + \epsilon_{22}  - \epsilon_{\beta} + \mathrm{E_{33}}  - \epsilon_{\gamma}]^2  \nonumber \\
& +\frac{\mathrm{C_{11}} - \mathrm{C_{12}}}{2}\bigg[(\mathrm{E_{11}}-\epsilon_{\alpha})^2 + (\mathrm{E_{22}}-\epsilon_{\beta})^2 + (\mathrm{E_{33}}-\epsilon_{\gamma})^2 \bigg]
\label{simple3}
\end{align}

Next taking plane stress conditions for the thin film, we have $\mathrm{E_{13}} = \mathrm{E_{23}} = 0$, and $\mathrm{E_{33}}$ is derived from Eq. \ref{eq:E33}. Furthermore, the mechanical constraints for the thin film conditions are given by $\mathrm{E_{11}} = \mathrm{E_{22}} = \mathrm{E_s}$ and $\mathrm{E_{12}} = 0$. We substitute these conditions in the above elastic energy expression:

\begin{align}
\psi_{elas}[c,\mathrm{E_s}] = \int_{{\mathcal{E}}}&\frac{\mathrm{C_{11}} - \mathrm{C_{12}}}{2}[(\mathrm{E_{11}}-\epsilon_{\alpha})^2 + (\mathrm{E_{22}}-\epsilon_{\beta})^2 + (\mathrm{E_{33}}-\epsilon_{\gamma})^2]\nonumber \\
& + \frac{\mathrm{C_{12}}}{2}[\mathrm{E_{11}} - \epsilon_{\alpha} + \mathrm{E_{22}}-\epsilon_{\beta} + \mathrm{E_{33}}-\epsilon_{\gamma}]^2\mathrm{d}\mathbf{x}.
\label{eq:EnergyIndicialNotationTF}
\end{align}

Substituting for film strains:
\begin{align}
\psi_{elas}[c,\mathrm{E_s}] = \int_{\mathcal{E}} &\frac{\mathrm{C_{11}}-\mathrm{C_{12}}}{2}\bigg[(\mathrm{E_s}-\epsilon_{\alpha})^2+(\mathrm{E_s}-\epsilon_{\beta}])^{2} + \bigg(\frac{\mathrm{C_{12}}}{\mathrm{C_{11}}}[\epsilon_{\alpha}+\epsilon_{\beta}-2\mathrm{E_{s}}]\bigg)^{2}\bigg]\nonumber \\
& +\frac{\mathrm{C_{12}}}{2}[\mathrm{E_s}-\epsilon_{\alpha} + \mathrm{E_s}-\epsilon_{\beta} + \frac{\mathrm{C_{12}}}{\mathrm{C_{11}}}(\epsilon_{\alpha}+\epsilon_{\beta}-2\mathrm{E_{s}})]^2\mathrm{d}\mathbf{x}.
\label{final}
\end{align}

\subsection*{Analytical calculation of in-plane stresses}
We define stresses as $\mathbf{\sigma}=\frac{\partial\Psi}{\partial\mathbf{E}}$, and assuming a linear elastic cubic relation, we have:
\begin{align}
\sigma_{11}=\mathrm{C_{11}}[\mathrm{E_{11}-\epsilon_{\alpha}}]+\mathrm{C_{12}}[\mathrm{E_{33}-\epsilon_{\gamma}}+\mathrm{E_{22}-\epsilon_{\beta}}]\nonumber\\
\sigma_{22}=\mathrm{C_{11}}[\mathrm{E_{22}-\epsilon_{\beta}}]+\mathrm{C_{12}}[\mathrm{E_{33}-\epsilon_{\gamma}}+\mathrm{E_{11}-\epsilon_{\alpha}}]\nonumber\\
\sigma_{33}=\mathrm{C_{11}}[\mathrm{E_{33}-\epsilon_{\gamma}}]+\mathrm{C_{12}}[\mathrm{E_{11}-\epsilon_{\alpha}}+\mathrm{E_{22}-\epsilon_{\beta}}]\label{eq:stress-relations}
\end{align}

Next, we substitute $\mathrm{E_{11}=E_{s}}$, $\mathrm{E_{22}=E_{s}}$, and from Eq.~\ref{eq:E33} $\mathrm{E_{33}=\frac{\mathrm{C_{12}}}{\mathrm{C_{11}}}}[(\epsilon_{\alpha}+\epsilon_{\beta}-2E_{s}]+\epsilon_{\gamma}.$ Note, the values of $\epsilon_{\alpha},\epsilon_{\beta},\epsilon_{\gamma}$ change as a function of Li-content($c$). Consequently the in-plane stresses evolve as a function of Li-content as:
\begin{align}
\sigma_{11} & =\mathrm{C_{11}}\mathrm{[E_{s}-\epsilon_{\alpha}}]+\mathrm{\frac{C_{12}^2}{C_{11}}}[\epsilon_{\alpha}+\epsilon_{\beta}-2\mathrm{E_{s}}]+\mathrm{C_{12}}[\mathrm{E_{s}-\epsilon_{\beta}}]\nonumber\\
 & =(\mathrm{C_{11}+C_{12}-2\frac{C_{12}^{2}}{C_{11}}})\mathrm{E_{s}+(-C_{11}}+\mathrm{\frac{C_{12}^{2}}{C_{11}})\epsilon_{\alpha}}+(\mathrm{-C_{12}+\frac{C_{12}^{2}}{C_{11}}})\epsilon_{\beta}\nonumber\\
\sigma_{22} & =\mathrm{C_{11}}[\mathrm{E_{s}-\epsilon_{\beta}}]+\mathrm{\frac{C_{12}^2}{C_{11}}}[\epsilon_{\alpha}+\epsilon_{\beta}-2\mathrm{E_{s}}]+\mathrm{C_{12}}[\mathrm{E_{s}-\epsilon_{\alpha}}]\nonumber\\
 & =\mathrm{(C_{11}+C_{12}-2\frac{C_{12}^{2}}{C_{11}})}\mathrm{E_{s}+\mathrm{(-C_{12}}+\frac{C_{12}^{2}}{C_{11}})\epsilon_{\alpha}}+(\mathrm{-C_{11}+\frac{C_{12}^{2}}{C_{11}})\epsilon_{\beta}}\nonumber\\
\sigma_{33} & =\mathrm{C_{12}}[\epsilon_{\alpha}+\epsilon_{\beta}-2\mathrm{E_{s}}]-\mathrm{C_{11}}[\epsilon_{\alpha}+\epsilon_{\beta}-\mathrm{2E_{s}}]\nonumber\\
 & =0 \label{eq:stresses-substrate-strain}
\end{align}

The analytical stresses in Eq.~\ref{eq:stresses-substrate-strain} are homogeneous in-plane stresses that evolve as a function of Li-content in $\mathrm{Li_xV_2O_5}$. These analytical expressions could be applied in a computational framework in which a spatially resolved Li-content $c(\mathbf{x})$ (i.e., composition microstructure) would be used to predict heterogeneous stress distribution in the thin-film $\mathrm{Li_xV_2O_5}$. We propose to study the heterogeneous stress distribution in a future work. For current purposes we analytically model the in-plane stress evolution during phase transformations and compare it with the stresses measured experimentally using the MOSS technique.

\subsection{Thin-film experiment}\label{Thin-film experiment}

In this section we give details of the thin-film preparation, characterization, electrochemical cycling and measurements. 

\subsubsection*{Film Preparation}
Fig.~\ref{fig:Schematic-illustration-of-thin-film}(a) shows a schematic illustration of the thin-film electrode
we prepared for the experiment. We deposited $\mathrm{50nm}$ film of $\mathrm{V_{2}O_{5}}$ electrode on two different substrates, namely,
an Aluminium-coated-quartz substrate and a Gold-coated-quartz substrate. The substrates were $\mathrm{250\mu m}$ thick and $25.4\mathrm{mm}$ in diameter. The substrates were thoroughly cleaned in acetone, ethanol,
and deionized water, and were mounted in an electron-beam evaporation system in order to deposit the metallic current collectors (Al or
Au). For the Au specimen, 10nm of Titanium was first deposited on quartz ($\mathrm{SiO_{2}}$) to improve adhesion between the Gold
and \textcolor{black}{quartz} substrate. Finally, the deposited $\mathrm{V_{2}O_{5}}$ layers were annealed at $550^{\circ}\mathrm{C}$ in the air for three
hours, in order to crystallize and oxidize them. Further details in preparing the thin-film substrate are provided by \cite{Sheth2017}.

\subsubsection*{Film characterization}
\begin{figure}[H]
\begin{centering}
\includegraphics[width=\textwidth]{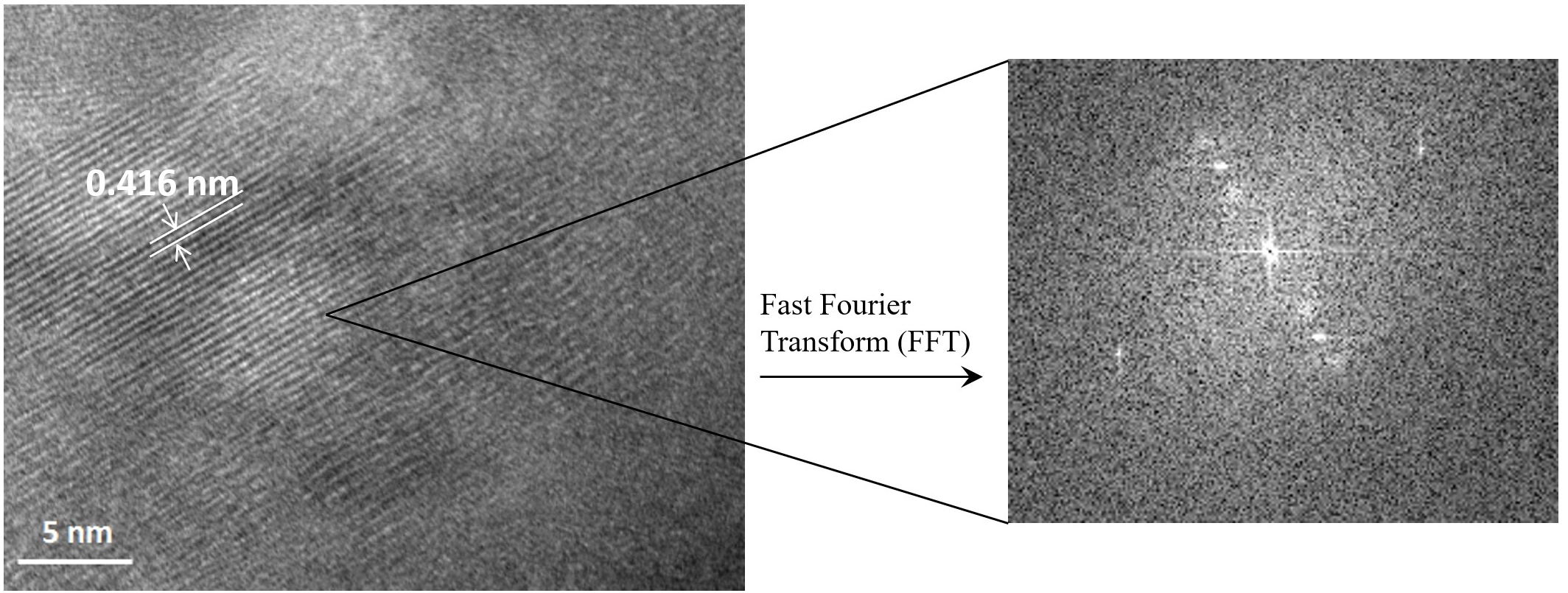}
\par\end{centering}
\caption{TEM image of the $\mathrm{V_2O_5}$ thin-film electrode on Au substrate showing its lattice fringes and the corresponding fast Fourier transformation of the TEM image confirms the crystallinity of $\mathrm{V_2O_5}$ before electrochemical cycling.\label{fig:TEM-image}}
\end{figure}

We characterized the thin films in two ways: First, we used scanning electron microscopy (SEM) to characterize the microstructures of the
thin-film electrodes. Figs.~\ref{fig:potentiostatic-cycling}(a-b) shows the SEM images of the $\mathrm{V_{2}O_{5}}$
surface on Al and Au substrates before electrochemical cycling. $\mathrm{V_{2}O_{5}}$
films on both Al and Au substrates are crystalline in the initial state. The crystallinity of these films was also verified by characterizing them using transmission electron microscopy (TEM), see Fig.~\ref{fig:TEM-image}. Second, we carried out X-ray diffraction (XRD) with a diffractometer using Cu $K_{\alpha}$ radiation ($\lambda=1.5406\angstrom$)
at glancing angles with the films. Figs.~\ref{fig:potentiostatic-cycling}(c) show the XRD measurements of $\mathrm{V_{2}O_{5}}$ films deposited on Al and Au substrates respectively. In the initial state, $\mathrm{V_{2}O_{5}}$ films on both Al and Au substrates crystallize into the orthorhombic structure, with a dominant peak corresponding to the $\mathrm{V_{2}O_{5}}(001)$ crystallographic direction. Other peaks correspond to the current collectors (Al or Au). The presence of a single dominant peak for
$\mathrm{V_{2}O_{5}}(001)$ suggests a polycrystalline texture of the thin-film electrode. Note, there is a bump in the XRD measurement on the Al substrate---we suspect that this arises from the quartz substrate present beneath the Al layer. In the case of Au, an in-between
Ti layer shields the XRD peak corresponding to the quartz substrate. Further details and specifications of the instruments used in the thin-film characterization is provided by \cite{Sheth2017}.

\subsubsection*{Electrochemical Cycling}
We designed four custom-made electrochemical
cells (labelled A,B,C,D) consisting of $\mathrm{V_{2}O_{5}}$ thin-film cathode, Li-metal
anode, and a liquid electrolyte (equimolar mixture of Ethylene Carbonate, Dimethyl Carbonate and 1M $\mathrm{LiPF_{6}}$ salt. In two of these cells (labelled A,B), the $\mathrm{V_{2}O_{5}}$ was deposited on Al-coated-quartz substrate, and in the other two cells (labeled C,D) $\mathrm{V_{2}O_{5}}$ was deposited on Au-coated-quartz substrate. We divided the four electrochemical cells into two studies: (a) Galvanostatic experiment: electrochemical cells A and C were charged/discharged under constant current density ($\mathrm{0.2\mu A/cm^{2}}$) and at a rate of C/20, and (b) Potentiostatic experiment: electrochemical cells B and D were cycled under constant voltage (defined here as current $<4\times10^{-8}\mathrm{A})$. Next, we give an overview of the electrochemical and stress measurements
made in-situ in the two studies.

\subsubsection*{Electrochemical and Stress Measurements}
\textit{Voltage measurement}: During each cycle, the electrochemical response (namely the voltage) was recorded. Fig.~\ref{fig:Galvanostatic-Cycling} shows the voltage plateaus for $\mathrm{V_{2}O_{5}}$ thin-film electrodes deposited on Al and Au current collectors, respectively, and cycled galvanostatically. These voltage plateaus resemble the characteristic features of the voltage curve for $\mathrm{V_{2}O_{5}}$ electrode measured in previous works \citep{christensen2018structural}, however, the exact voltage values at which these plateaus appear are shifted to lower values. As predicted the film strain imposed on the thin-film electrode plays a key role in shifting the voltage curves and we discuss this in detail in Sections 2 and 3.\vspace{2mm}

\textit{Stress measurement}: During each cycle, the in-plane stresses generated during Li insertion and extraction were measured in-situ using the multi-beam optical stress sensor (MOSS) technique. This technique
employs an array of parallel laser beams that are focussed on the underside of the quartz substrate and help monitor the changes in substrate curvature during electrochemical cycling. The substrate curvature $\kappa$ is used as a measure to estimate average in-plane stresses $<\sigma>$ in the thin-film using Stoney's equation, $<\sigma>h_{\mathrm{film}}=\frac{M_{\mathrm{sub}}h_{\mathrm{sub}}^{2}\kappa}{6}.$ Here, $h_{\mathrm{film}}$ and $h_{\mathrm{sub}}$ are the thickness of the film and substrate respectively, and $M_{\mathrm{sub}}$ is the biaxial modulus of the substrate. Specific details of measuring stresses
using the MOSS technique are described by \cite{Sheth2017,sethuraman2010situ,tripuraneni2018situ}. The in-plane stresses measured during the Galvanostatic experiment are shown in blue in Fig.~\ref{fig:Stress-Evolution}. These stresses are sensitive to the chemo-mechanical strains in $\mathrm{V_{2}O_{5}}$ electrode during
electrochemical cycling and evolve as a function of Li-content.\vspace{2mm}

\textit{XRD and SEM measurements}: We conducted post-cycling XRD measurements after three charge/\\discharge potentiostatic cycles of the $\mathrm{V_{2}O_{5}}$ electrode. The corresponding XRD peaks are shown in Fig.~\ref{fig:potentiostatic-cycling}(c). For $\mathrm{V_{2}O_{5}}$ deposited on Al-coated-quartz substrate the diffraction peak corresponding to $\mathrm{V_{2}O_{5}}(001)$ broadens with repeated cycling but does not disappear. This shows that $\mathrm{V_{2}O_{5}}$ undergoes structural transformations during lithiation/delithiation, but retains its crystallinity upon repeated cycling. However, for $\mathrm{V_{2}O_{5}}$ deposited on Au-coated-quartz substrate $\mathrm{V_{2}O_{5}}(001)$ peak
disappears after 3 cycled indicating a loss of crystallinity (or amorphization) in the $\mathrm{V_{2}O_{5}}$ thin-film. This was further confirmed by SEM images that show a powdery microstructure after 3 cycles on $\mathrm{V_{2}O_{5}}$ deposited on gold substrate, see Fig.~\ref{fig:potentiostatic-cycling}(b).

\newpage
\bibliography{reference}
\bibliographystyle{elsarticle-harv}
\setcitestyle{authoryear,open={(},close={)}}
\end{document}